\documentclass[12pt,preprint]{aastex}

\newcommand{\arcm}{\ifmmode {' }\else $' $\fi}
\newcommand{\arcs}{\ifmmode {'' }\else $'' $\fi}
\newcommand{\kms}{km\thinspace s$^{-1}$}
\slugcomment{}

\shortauthors{Bridges et al.} \shorttitle
{Spectroscopy of Globular Clusters in M104}

\begin{document}

\title{Spectroscopy of Globular Clusters out to Large Radius in the 
Sombrero Galaxy}

\author{Terry J. Bridges}
\affil{Department of Physics, Queen's University, Kingston, ON
 K7L~3N6, Canada; tjb@astro.queensu.ca}

\author{Katherine L. Rhode\altaffilmark{1}}
\affil{Astronomy Department, Wesleyan University, Middletown, CT
  06459; kathy@astro.wesleyan.edu\\
and\\
Department of Astronomy, Yale University, New
  Haven, CT 06520}
%\email{rhode@astro.yale.edu}

\author{Stephen E. Zepf}
\affil{Department of Physics \& Astronomy, Michigan State University,
  East Lansing, MI 48824; zepf@pa.msu.edu}

\author{Ken C. Freeman}
\affil{Research School of Astronomy \& Astrophysics, Australian
  National University, Mount Stromlo Observatory, Weston
  Creek, ACT 2611, Australia; kcf@mso.anu.edu.au}

\altaffiltext{1}{NSF Astronomy \& Astrophysics Postdoctoral Fellow}

\begin{abstract}
We present new velocities for 62 globular clusters in 
M104 (NGC 4594, the Sombrero Galaxy), 56 from 2dF on the AAT and 6 from
Hydra on WIYN. Combined with previous data, we have a total sample
of 108 M104 globular cluster velocities, extending to 
20\arcmin\ radius ($\sim$ 60 kpc), along with BVR photometry
for each of these.
We use this wide-field dataset to study the globular cluster kinematics
and dark matter content of M104 out to 10\arcmin\ radius (30 kpc).
We find no rotation in the globular cluster system. The edge-on nature
of M104 makes it unlikely that there is strong rotation which is
face-on and hence unobserved; thus, the absence of rotation over our 
large radial range appears to be an intrinsic feature of the 
globular cluster system in M104.
We discuss ways to explain this low rotation, including the 
possibility that angular momentum has been transferred to even 
larger radii through galaxy mergers.
The cluster velocity dispersion is $\sim$ 230 km/s within several 
arcmin of the galaxy center, and drops to $\sim$ 150 km/s at 
$\sim$ 10\arcmin\ radius. We derive the mass profile of M104
using our velocity dispersion profile, together with
the Jeans equation under the assumptions of spherical symmetry
and isotropy, and find excellent agreement with the mass inferred from
the stellar and gas rotation curve within
3\arcmin\ radius. The M/L$_V$ increases from $\sim$ 4 near the galaxy
center to $\sim$ 17 at 7\arcmin\ radius ($\sim$ 20 kpc, or 4 R$_e$), 
thus giving strong support for the presence of a dark matter halo in
M104. 
More globular cluster
velocities at larger radii are needed
to further study the low rotation in the globular
cluster system, and to see if the dark 
matter halo in M104 extends beyond a radius of 30 kpc.

\end{abstract}


\section{Introduction}

The prevailing view of galaxy formation is that galaxies assemble
hierarchically from smaller structures that are composed of both dark
and baryonic matter.  These structures collide and merge to create
larger structures, with the eventual result being a bound galaxy in
which the luminous, baryonic matter exists within a much more massive
halo of dark matter.
Testing this paradigm is crucial to our developing a complete,
self-consistent picture of cosmology and galaxy formation.

Although one can in theory measure the masses and mass profiles of
galaxies using a variety of methods --- e.g., observations of
integrated starlight, HI in late-type galaxies, and X-ray-emitting gas
in luminous ellipticals --- in practice this can be difficult,
particularly for early-type galaxies. The challenges for early-type
galaxies are that they normally lack significant amounts of extended
HI gas, many do not have luminous, extended hot gaseous halos for
X-ray studies, and their integrated starlight can only be measured to a
few effective radii (e.g. Kronawitter et al. 2000).
Globular clusters (GCs) are luminous, compact objects that are distributed
more or less spherically around giant galaxies, number in the hundreds
to thousands, and are readily detected in wide-field imaging out to
10$-$15 $R_{e}$ (e.g. Rhode \& Zepf 2001, 2004), 
and for these reasons make
uniquely valuable tracers of galaxy structure.  Furthermore, they may
be less kinematically biased than other types of dynamical tracers;
for instance, the numerical simulations of Dekel et al. (2005) show that
planetary nebulae in early-type galaxies may be on very elongated, radial
orbits as a result of disk-galaxy mergers.
Since most GCs are
old and are likely markers of the major star formation episodes that a
galaxy has undergone (e.g., Ashman \& Zepf 1998, Brodie \& Strader
2006), they also provide an observable record of the formation and
assembly history of galaxies.  More specifically, many galaxies have
been found to have two populations of GCs: a blue, metal-poor
population and a red, more metal-rich one, that appear to have formed
in different episodes (e.g. Gebhardt \& Kissler-Patig 1999; 
Kundu \& Whitmore 2001). Galaxy formation models that predict how these
populations arose in the context of a galaxy's formation often predict
that the red and blue populations will have different kinematics.
Measuring GC velocities therefore provides a test of the proposed
formation scenarios. 

To date, only a small sample of galaxies have had substantial numbers
($\sim$100 or more) of their GC velocities measured.
Three of these --- M87 (Cohen 2000, C\^{o}t\'{e} et al. 2001), NGC~4472 (Zepf
et al.\ 2000; C\^{o}t\'{e} et al. 2003), and NGC~1399 (Richtler et al.\ 2004)
--- are luminous ellipticals located near the centers of galaxy
clusters, one (NGC~5128; Peng et al.\ 2004) is a moderate-luminosity
elliptical with a peculiar morphology (possibly due to a recent merger)
and two are spirals --- our own galaxy (see Harris 1996 for a
compilation) and M31 (e.g., Perrett et al.\ 2002).  Studying GC
kinematics in galaxies over a wider range of luminosities and
environments is necessary before we can begin to draw general
conclusions about galaxy formation
and how galaxy mass profiles change with overall galaxian properties.
We also need to measure GC velocities over a larger radial range than
typically has been done in past studies, which have for the most part
concentrated on the central regions of galaxy GC systems.  Covering a
large radial range is especially important for quantifying the
distribution of dark matter in galaxy halos, and for studying GC
kinematics at large radius. 

The Sombrero galaxy (NGC~4594, M104) is an interesting target for a GC
spectroscopic study because it is the closest 
undisturbed field galaxy with a
luminous bulge/spheroid. M104 has M$_V$ = $-$22.4, typical of
giant ellipticals, and is intermediate in its properties between
spiral and elliptical galaxies. Table~\ref{table:galaxy properties}
summarizes some of these properties.  It is sometimes classified as an Sa
spiral (e.g., de~Vaucouleurs et al.\ 1991), but its large bulge-to-disk
ratio and bulge fraction are more like that of an S0 (Kent 1988), and
its optical colors are likewise similar to those of S0s (Roberts \&
Haynes 1994). M104 has two advantages, however, over giant elliptical
galaxies. First, measurement of its disk rotation out to $\sim$ 3\arcmin\
gives an independent constraint on the mass profile out to moderate radii
(see Section \ref{dmhalo}). Second, since M104 is reasonably isolated, GC
kinematics probe only its gravitational potential, and not that of a
surrounding galaxy group or cluster. 

The Sombrero is relatively nearby (9.8~Mpc; Tonry et
al. 2001) and its GC system has been studied with photographic plates
(e.g., Harris et al.\ 1984), CCD detectors (Bridges \& Hanes 1992,
Rhode \& Zepf 2004; hereafter RZ04), and
Hubble Space Telescope imaging (Larsen et al.\ 2001, Spitler et al.\
2006). RZ04 imaged the galaxy out to a radius of $\sim$65~kpc with a
mosaic CCD detector and multiple broadband filters, and used these data
to derive global properties for the GC system.  Selecting GC
candidates in multiple filters reduced the contamination from
foreground and background objects, although contamination from stars
remains significant because of the Sombrero's location toward the
Galactic bulge.  RZ04 found that M104 has $\sim$1900 GCs, a spatial 
extent of $\sim$50~kpc, and a specific frequency S$_N$
(GC number normalized by the $V$-band luminosity of the galaxy, as
defined by Harris \& van den Bergh 1981) of 2.1$\pm$0.3.  The color
distribution of the system is bimodal, with 
about 60\% blue (metal-poor) GCs
and 40\% red (metal-rich).  The blue GC population is slightly more
extended than the red population, producing a shallow color gradient
in the overall system.

\begin{deluxetable}{ccccccccccc}
%\rotate
\tablecaption{Properties of the Sombrero and Its Globular Cluster System}
\tablewidth{490pt}
\tablehead{
\multicolumn{6}{c}{General Properties} & \colhead{} & \multicolumn{4}{c}
{GC System Properties}\\
\cline{1-6} \cline{8-11}\\
\colhead{$B/T$} &\colhead{$R_e$} 
&\colhead{Dist} &\colhead{E(B-V)} &\colhead{$V_T^0$} &\colhead{$M_V^T$} &\colhead{}
&\colhead{Extent}&\colhead{$S_N$}&\colhead{N(GC)} & \colhead{Blue/Red}\\
\colhead{} & \colhead{(\arcsec/kpc)} & \colhead{(Mpc)} & \colhead{} & \colhead{} & \colhead{}
& \colhead{} & \colhead{(kpc)}& \colhead{} & \colhead{} & \colhead{(\%)} }
\startdata
%Morph Type  B/T   B-V  m-M   Dist   M_V  GC S_N  N_GC  GCS_ext  blue/red
%
0.86 & 105/5 & 9.8 & 0.051 & 7.55 & $-$22.4 & & 50 & 
2.1$\pm$0.3 & 1900 & 60/40 \\
\enddata

\tablecomments{B/T is from Kent (1988); $V_T^0$ is from RC3
(deVaucouleurs et al.\ 1991). Distance is from Tonry et al.\
(2001). E(B$-$V) is from Schlegel et al. (1998). 
$M_V^T$ is from combining $V_T^0$ with distance. 
Effective radius R$_e$ is from Burkhead (1986).
GC system properties are from RZ04.}  

\protect\label{table:galaxy properties}
\end{deluxetable}

Spectroscopy of GCs in M104 has been published in two previous
studies.  Bridges et al.\ (1997; hereafter B97) used the William
Herschel Telescope (WHT) to measure radial velocities of 34 GCs out
to 5.5$\arcm$ ($\sim$16~kpc) from the galaxy's center, with velocity
errors of 50$-$100~\kms. From these velocities they estimated a mass
of 5$^{+1.7}_{-1.5} \times$10$^{11}$ M$_\odot$ for M104, and found
that the M/L increases with radius, in other words that M104 possesses
a dark matter halo.
The second spectroscopic
study was done by Larsen et al.\ (2002; hereafter L02), who measured
spectra of 14 GCs in M104 with the Keck~I telescope.
The GCs in the L02 study are located within 5$\arcm$ of the galaxy
center, with nearly all (80\%) of them in the central 2$\arcm$.
L02 estimated the galaxy's mass within this central region, and also
combined their velocities with those of B97 to determine a 
projected mass of (5.3$\pm$1.0)$\times$10$^{11}$M$_\odot$ within
17 kpc.

In this paper, we present the results from spectroscopy
of 62 GCs in M104.  Fifty-six of the GCs were observed with the
3.9-m Anglo-Australian Telescope (AAT) and 2dF multi-fiber spectrograph.
Six more GC spectra were obtained with the 3.5-m WIYN telescope and
the Hydra fiber positioner and bench spectrograph\footnote{The WIYN
Observatory is a joint facility of the University of Wisconsin,
Indiana University, Yale University, and the National Optical
Astronomy Observatories}.  The target objects were
identified in the mosaic CCD survey of RZ04 and are located between 2
and 20$\arcm$ from the galaxy center.  The data presented here double
the number of known GC velocities for this galaxy and, combined with
data from B97 and L02, bring M104 into a sample of only seven galaxies
with $>$100 measured GC velocities. 
Furthermore, this study increases the radial coverage for M104 by
nearly a factor of four compared to the previous studies.
This enables us for the first time to probe the kinematics of M104's
outer halo and GC system, and to trace the galaxy's mass distribution
to many effective radii.

In the following section, we describe the observations and the steps used
to reduce and analyze the data. In Section \ref{id} we present our
sample of new radial velocity measurements for 62 M104 GCs, which in 
combination with previous data
yields a total sample of 108 GC radial velocities in M104. 
In Section \ref{results} we
present and discuss 
our results, including an analysis of the kinematics of the GC
system, the GC velocity dispersion profile, and the mass profile of M104. 
Finally, in Section \ref{conclusions}, we 
summarize the main results of this study. 
Throughout this paper, we adopt a distance of 9.8 Mpc for 
M104 (Tonry et al. 2001), and 
an effective radius R$_e$ = 105\arcsec\ (5 kpc for our 
adopted distance) (Burkhead 1986).

\section{Observations, Reductions, \& Analysis}
\label{data}

\subsection{Target Selection \& Observations}

To create a list of targets for this study, we began with a
preliminary list of 
GC candidates produced from $BVR$ images of M104 taken with the Mosaic
Imager on the Kitt Peak National Observatory 4-m Mayall telescope.
The final results from the 4m-Mosaic survey of M104's GC system are
published in RZ04. The survey techniques are detailed there; briefly,
objects qualify as GC candidates if they appear as point sources in
the 36$\arcm$ x 36$\arcm$ Mosaic images, are detected in all three
filters, and have $BVR$ magnitudes and colors consistent with what one
would expect for GCs at the distance of the galaxy (see RZ04 for
details of the selection methods). RZ04 identified 1748 unresolved GC
candidates in M104; the final set of candidates
have $V$ magnitudes between 18.96 and 24.3, $B-V$
colors in the range 0.32$-$1.24 and $V-R$ colors between 0.23 and 0.78.

\subsubsection{AAT/2dF Targets and Observations}

A preliminary photometric calibration of the Mosaic data was done in
2001 and a list of $\sim$1900 GC candidates was produced.  (The
photometric calibration was later redone using new data, and revised
magnitudes and colors were calculated for all the Mosaic sources.
Some of the original GC candidates were rejected based on the revised
photometry; the final list of GC candidates includes the 1748 objects
mentioned above.)
Starting with this list, we selected a subset of
584 objects with 19.0 $<$ $V$ $<$ 21.5. 
The Mosaic images were calibrated astrometrically using tasks in the
IRAF\footnote{IRAF is distributed by the National Optical Astronomy
Observatories, which are operated by the Association of Universities
for Research in Astronomy, Inc., under cooperative agreement with the
National Science Foundation.}  IMCOORDS package and coordinates for
stars in the USNO-A2.0
Catalog (Monet et al. 1998). The astrometric solution has an
rms of $\le$ 0.4\arcsec, and this accuracy has been confirmed 
by matching Chandra sources with several of the RZ04 object positions.
This input list was then weighted by
magnitude and radius, with bright candidates at large radius
given the highest weight.  

199 of these 584 GC candidates were observed with the 2dF multi-fiber
spectrograph on the AAT in April 2002.
The 2dF instrument has 400 fibers over a two-degree field of view
(FOV), making it well-suited to wide-field GC spectroscopy (Lewis et
al. 2002). The 2dF Configure software was used to automatically select
these 199 objects; this pointing also included 78 fibers positioned on
blank sky, and 4 fiducial fibers for field acquisition and
guiding. 2dF has two spectrographs, with each spectrograph receiving
200 fibers. We used 600V gratings in both spectrographs, centered at
5000 \AA, with spectral coverage from 3900$-$6100 \AA.  The dispersion
is 2.2 \AA/pixel, and the resolution was 4.5 and 5.5 \AA\ for the two
spectrographs (spectrograph \#1 has poorer resolution). The 2dF fiber
size varies between 2$-$2.1\arcsec\ across the field.

Our observing sequence consisted of a fiber flatfield at the
beginning, followed by 1800 sec object exposures; CuAr+CuHe arcs were
taken after every two object exposures. For each sequence we also
obtained 3$\times$300 sec offset sky exposures, where the telescope is
offset a few arcmin from the field; in the end, however, we did not
use these for sky subtraction (see Section~\ref{section:2dF redux}). On
17 April 2002, we obtained 2$\times$1800 sec object exposures under
poor conditions (seeing ranging from 2.4$-$3\arcsec). On 18 April,
conditions were better (some haze, seeing starting at 2.0\arcsec,
improving to 1.5$-$1.8\arcsec\ through the night), and we obtained
14$\times$1800 sec object exposures. Thus, we obtained a total of 8
hours on-source over the two nights. We also observed six radial
velocity standard stars: HD043318 (F6V), HD140283 (sdF3), 
HD157089 (F9V), HD165760 (G8III), HD176047 (K0III), and HD188512 (G8IV), 
and one flux standard star (EG274) through one fiber in each spectrograph. 

\subsubsection{WIYN/Hydra Targets and Observations}

To select targets for WIYN/Hydra, we began with the final list of 1748
GC candidates from RZ04.  From this list, we chose objects without
measured radial velocities and with $V$ magnitudes between 19.5 and
20.8.  The bright-end limit was imposed to reduce contamination from
Galactic stars; our 2dF results indicated that the rate of stellar
contamination is substantially higher for GC candidates with $V$
between 19.0 and 19.5.  We also included a few GCs that we had
observed with 2dF, in order to check the agreement between the 2dF and
Hydra velocities.

Hydra currently has $\sim$80 fibers that can be positioned over a
1-degree field of view.  We created three Hydra configurations
but were able to observe only one of these.  The pointing we observed
included 51 GC candidates, five GCs (with $V$ $=$ 19.2$-$19.4)
previously observed with 2dF, 15 fibers positioned on blank sky, and
six fibers positioned on guide stars.
We used the Bench Camera, red fiber cable, and 600@10.1 grating.  The
spectra were centered at 5300\AA\ and covered the region
3900$-$6800\AA, with a dispersion of 1.4\AA\ per pixel.  The spectral
resolution, given the typical FWHM of the slit profile of 2.5 pixels,
is 3.5\AA.

We were scheduled on WIYN for three nights in March 2006, but were
only able to take data on the night of 26 March 2006.  The rest of the
time was lost due to mechanical problems and bad weather.  On 26
March, we obtained five 2400-second exposures (total integration 3.3
hours) of the above-mentioned pointing under clear conditions. We had
planned to spend at least 6$-$8 hours observing this pointing, so only
the brightest targets in the configuration had enough signal-to-noise
to yield reliable cross-correlations 
(see Section \ref{section:measure_velocities}). We took a
series of dome flats and CuAr comparison lamp observations before and
after the object frames.  That same night we also observed three
radial-velocity standard stars (through a single fiber) for use as
cross-correlation templates: HD~65934 (G8III), HD~86801 (G0V), and 
HD~90861 (K2III).

\subsection{Spectroscopic Data Reduction}

\subsubsection{2dF Data}
\label{section:2dF redux}

We used the 2dfdr software package (Croom et al. 2005) to reduce the AAT 2dF
data. The reduction sequence was as follows.
Bias subtraction was done in the standard way using the overscan
region. Fiber flatfields were used to trace the spectra on the 
CCD, also called ``tramline mapping''. 
Fiber extraction was done using the ``FIT'' algorithm, which
performs an optimal extraction based on the fitting profiles determined
from the fiber flatfield in the previous step.
Wavelength calibration was done using the CuAr+CuHe arcs, with
typical rms of 0.15-0.2 \AA.
Before sky subtraction is done, one needs to correct for
fiber-to-fiber differences in throughput, and to normalize all 
fibers to the same level. This was done using the ``skyflux'' 
method, where the flux in night sky lines is used to determine
the relative fiber throughput.
Sky subtraction was done by taking the median of the 
normalized sky fibers to form a combined sky spectrum, which is
then subtracted from each fiber spectrum (both object and sky
fibers). The sky subtraction accuracy is defined as the fraction of
residual light in the sky fibers after sky subtraction, taken as 
a mean over all sky fibers. Our sky subtraction accuracy varied
between 2$-$4\% over our 16 frames.
Finally, reduced frames were combined with optimal S/N and
with cosmic ray rejection (this 2dfdr algorithm is based on the
IRAF imcombine/crreject algorithm). Flux weighting was done using
the flux in the brightest 5\% of fibers to weight each frame.

The 2dF spectra of radial velocity and flux standard stars were
reduced in the same manner, except that throughput calibration and sky
subtraction were not necessary for these short exposures
(typically a few sec).

\begin{figure}
\epsscale{0.75}
\plotone{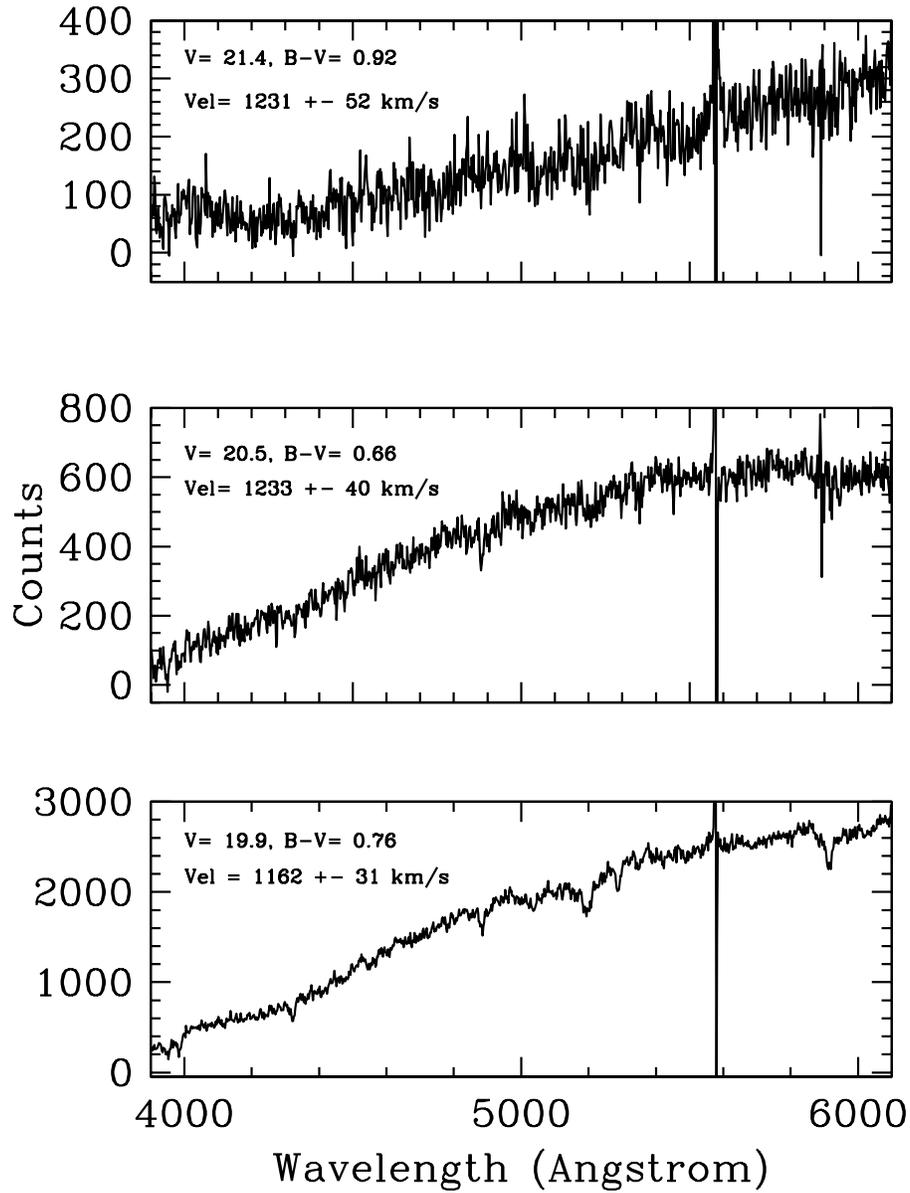}
\caption{Representative 2dF spectra of M104 GCs, from low S/N (top),
intermediate S/N (middle), and high S/N (bottom).}
\label{fig:sample_spectra}
\end{figure}

Figure \ref{fig:sample_spectra} shows some illustrative 2dF spectra of
varying S/N.

\subsubsection{Hydra Data}

Initial reduction of the Hydra data was done with standard IRAF tasks:
ZEROCOMBINE to construct a combined bias frame; CCDPROC to
bias-subtract and trim the target images, flats, and CuAr comparison
lamp images; and FLATCOMBINE to create stacked dome flats.  The IRAF
task DOHYDRA was then used to extract the spectra and perform
throughput correction, flat-fielding, wavelength calibration, and sky
subtraction of the target exposures. The sky subtraction was
accomplished by examining the individual spectra in each sky fiber,
rejecting those that appeared contaminated by a nearby source, and
averaging the remaining spectra using a sigma-clipping algorithm for
cosmic-ray removal.  The same steps used for the target object images were
also used to reduce the three standard star spectra.  

The five 2400-sec integrations of the target field were scaled to the
same flux level and then combined (with cosmic-ray rejection) using
the SCOMBINE task.  Approximately 50\AA\ was clipped from each of the blue and
red ends of the combined object spectra to remove low signal-to-noise
regions.  Finally, the continuum level was fit with a polynomial and
subtracted from each spectrum.  Clipping and continuum-subtraction
were also performed on the standard star spectra.

\subsection{Measuring Radial Velocities}
\label{section:measure_velocities}

Heliocentric radial velocities for the target objects were derived
using the IRAF task FXCOR, which performs Fourier cross-correlation of
an object spectrum against a specified template spectrum. For the
cross-correlation of the 2dF spectra, we used the radial velocity
standard stars HD043318, HD157089, HD16570, HD176047, and HD188512 as
templates, with heliocentric velocities obtained from Barbier-Brossat
et al.(1994), Barbier-Brossat \& Figon (2000) and 
Malaroda et al. (2001) via SIMBAD. We used the
wavelength region from 3900$-$6000 \AA\ for cross-correlation (masking
off a region around the night sky-line at 5577 \AA, where imperfect
sky subtraction can lead to spurious cross-correlations), and we
continuum-subtracted and ramp-filtered the spectra. We used only those
templates giving a Tonry-Davis R coefficient $>$ 2.5, and we demanded
that we have at least two templates with reliable velocities for a
given object spectrum (this last condition only removed one possible
GC).  The final velocities were obtained from an average of the
velocities from the five templates, weighted by the Tonry-Davis R
coefficient.

The Hydra target spectra were cross-correlated against the three IAU
radial velocity standard stars we observed with Hydra: HD65934,
HD86801, and HD90861.  Regions of the spectra around night sky lines
at 5577\AA, 5892\AA, 6300\AA, and 6364\AA\ were excluded from the
cross-correlation.  The final measured radial velocities were
calculated from the weighted mean of the velocities obtained from the
three templates.

\section{Globular Cluster Sample}
\label{id}

Of the 199 objects we observed with 2dF, 163 yielded reliable radial
velocities. An additional object is likely a QSO at z=1.3 (RZ \#1674, 
RA/Dec: 12:40:23.63/-11:41:04.0), while the
remaining 35
objects lacked sufficient signal-to-noise to measure their velocities.
Because the Hydra observations had a much shorter integration time
than planned, the spectra from Hydra had relatively low
signal-to-noise.  As a result only ten of the 51 objects we observed
with Hydra yielded reliable radial velocities. 

\begin{figure}
\epsscale{1.0}
\plotone{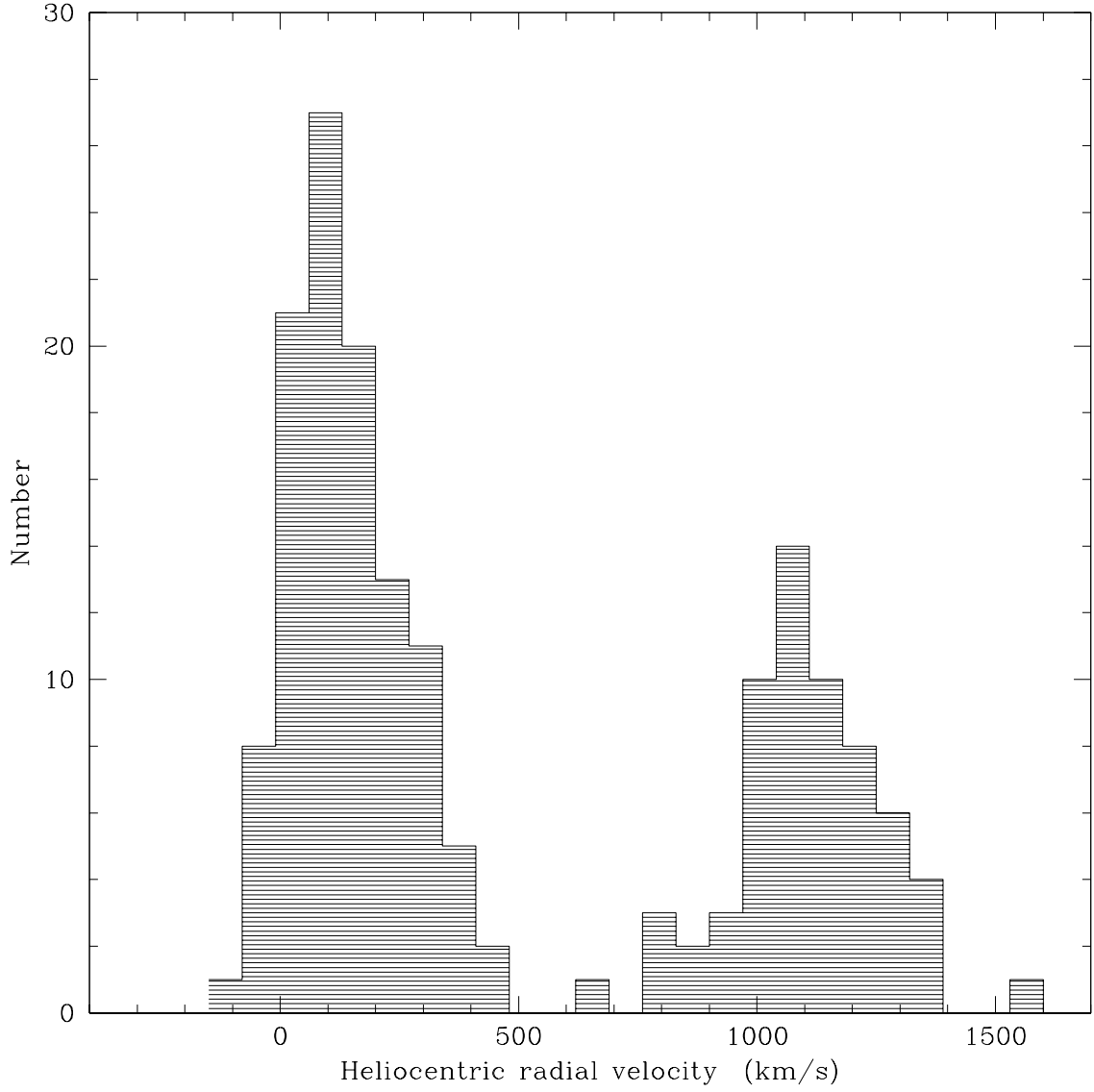}
\caption{Heliocentric radial velocity for 170 objects observed with 2dF
and Hydra.  Note the clear division between stars (with radial
velocity $<$500~km/s) and GCs in M104 (with radial velocity between
600 and 1600~km/s).}
\label{fig:vel_hist}
\end{figure}

In Figure \ref{fig:vel_hist} we show a histogram of the objects with
reliable velocities derived from the 2dF and Hydra data. 
There is a clear separation between objects with velocities $<$ 500
km/s, which are likely to be stars, and those with velocities between
600$-$1600 km/s, which are likely to be GCs in M104.  The
systemic radial velocity for M104 from RC3 (de~Vaucouleurs et al.\
1991) is 1091$\pm$5~km/s.
We adopt a velocity of 500 km/s as the division between GCs and
non-GCs.

Of the 163 objects with reliable radial velocities from 2dF, 
56 are {\em bona fide} GCs in M104, and 107 are stars. 
Of the ten objects from Hydra with measured radial velocities,
one is a star, three are repeat observations of bright GCs from the 2dF data,
and six are new GCs in M104.  The radial velocities of the three GCs
observed with both 2dF and Hydra agree within their errors, with
2dF/Hydra velocities of 818 $\pm$ 22/797 $\pm$ 40, 
1046 $\pm$ 24/1079 $\pm$ 37, and 1278 $\pm$ 28/1309 $\pm$ 79 km/s.
For these cases we have adopted the 2dF value for the radial velocity.

We note that the spectroscopic samples we chose have a very low rate
of contamination from background objects: only one background object
(RZ\#1674, the likely QSO) was found in the sample of 170 objects
for which we measured radial velocities.
However the sample has a high rate of contamination from
foreground stars. This can be attributed primarily to M104's location
in the sky, at fairly low Galactic latitude in the direction of the
bulge (l= 298, b= 51 deg).
RZ04 ran a model
code in order to estimate the number density of Galactic stars within
a given magnitude and color range in a given direction on the sky.
The result was that the predicted stellar contamination in the direction 
of M104
was at least a factor of 2$-$3 larger than the contamination for the
other galaxies we have surveyed.  Another reason for the high stellar 
contamination
is that very preliminary photometry was used to select GC candidates
for the 2dF observing run.  The measured magnitudes and colors of the
objects in the KPNO 4m/Mosaic images of M104 changed significantly
when the final photometric calibration was done (using
post-calibration data obtained at the WIYN telescope, after the 2dF
observing was completed).  The improved $BVR$ photometry eliminated
from our final GC candidate list a total of 36 of the 108 foreground stars,
as well as the QSO.
Finally, by observing
objects with $V$ $>$ 19.5 in future runs, we can further minimize
contamination from Galactic stars, since many of the GC candidates which
turned out to be stars have 19.0 $<$ $V$ $<$ 19.5.

We next combine our 56 2dF velocities with those from the previous
WHT study of B97, and the Keck study of L02. 
There are 34 confirmed GCs from B97, 
and 14 from L02 (we follow L02 in not using 
object H2-27, which has poor S/N and an uncertain velocity).
One of the WHT GCs (1-12) matches a
2dF GC, with the WHT and 2dF velocities being 1370 $\pm$ 32 km/s and
1350 $\pm$ 42 km/s respectively; we adopt the 2dF velocity. There is 
also another match between a WHT GC (2-8) and a Keck GC (C-134); 
the WHT and Keck velocities are 979 $\pm$ 14 km/s and 950 $\pm$ 15 km/s
respectively, and we adopt the WHT velocity. 
In addition, we have recent WIYN/Hydra velocities for 6 new M104 GCs,
as presented earlier.
We thus have a total of 56 + 6 + 33 + 13 = 108 independent M104 GC
velocities from the 2dF/Hydra/WHT/Keck data. The number of overlaps
is small, but the velocity differences are all less than 35 km/s, 
giving confidence that the datasets can be combined. 

Table \ref{tab_globs} lists identification numbers, 
positions, heliocentric radial velocities,
major/minor axis distances, galactocentric radii, azimuthal angles, and
$BVR$ photometry for the total sample of 108 GCs from 2dF, Hydra, WHT,
and Keck. The identification number given in column 1 is a sequence 
number assigned for the RZ04 study.  Column 2  gives (when applicable) 
the sequence number from B97 or L02. We adopt a position angle of 90 deg 
for the semimajor axis of M104 (RC3, Ford et al. 1996) and a galaxy center of 
12~39~59.43 -11~37~23.0 (J2000) (Petrov et al. 2006).  
$\theta$ = 0 corresponds to +X and East on the sky.  
The photometry for all objects was measured from the $BVR$ images from
the RZ04 mosaic CCD study.  19 of the GCs listed in
Table \ref{tab_globs} (mainly those from L02, plus a few from B97)
were not included in the list of GC candidates found by RZ04 because
they were located close to the galaxy center, in regions of high
galaxy background that had been excluded from the search for GC
candidates described in that paper. For the current study, we located
those sources in the RZ04 images and measured their $BVR$ magnitudes
in order to include them in the table. The result is that Table
\ref{tab_globs} presents a consistent, uniform set of $BVR$ colors
for our full sample of 108 GCs with measured radial velocities in M104.
RZ04 found that M104 has the bimodal GC colour distribution typical of 
most early-type galaxies, and by applying the KMM mixture modelling
algorithm found a separation between the blue/metal-poor and
red/metal-rich GCs at B-R=1.3. We adopt this split, and find that we
have 66 blue GCs (B-R $<$ 1.3) and 42 red GCs (B-R $>$ 1.3); our
percentage of blue GCs (61\%) is similar to that found by RZ04 for
the complete photometric sample (59-66\%).
Table \ref{tab_nonGCs} lists positions, velocities, and $BVR$ photometry
(again, from the RZ04 mosaic images)
for the foreground stars found from the 2dF and Hydra spectra.

\begin{figure}
\epsscale{1.0}
\plotone{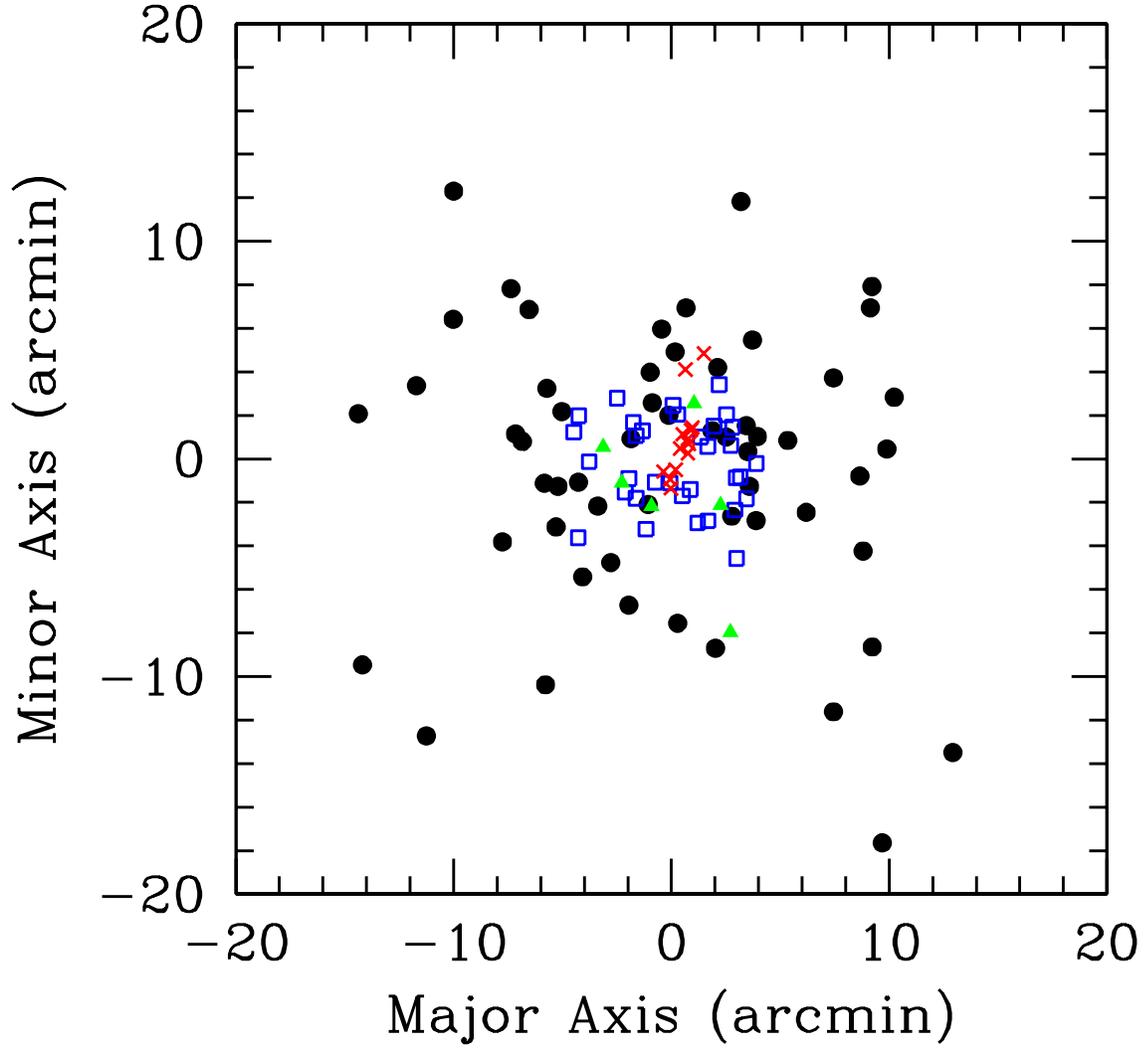}
\caption{Location of Confirmed M104 GCs. Filled (black) circles: 2dF; 
Open (blue) squares: WHT; (red)  crosses: Keck; Filled (green) 
triangles: Hydra.}
\label{fig:location_all}
\end{figure}

Figure \ref{fig:location_all} plots the GC major and minor axes to 
show the spatial coverage of the four datasets, and illustrates the 
expanded spatial coverage of our 2dF and Hydra data compared to previous WHT
and Keck data. For the systemic velocity of M104 in all subsequent analysis, 
we adopt the mean velocity of our GC sample, which is
1083 $\pm$ 20 km/s based on the biweight determinations using the
ROSTAT code (Beers et al. 1990).
Our GC velocity agrees well with the M104 velocity found by RC3, 
Rubin et al. (1978), and Faber et al. (1977), who measured
1091 $\pm$ 5, 1076 $\pm$ 10, and 1089 $\pm$ 15 km/s, respectively. The
velocity dispersion of our full M104 GC sample is 204 $\pm$ 16 km/s.
We discuss rotation of the GC system, and the radial profile of the 
GC dispersion and its implications in the following section.

\begin{deluxetable}{cccccccccccccccc}
\tabletypesize{\scriptsize}
\rotate
\tablecaption{Spectroscopically Confirmed M104 Globular Clusters.
Successive columns give: ID from RZ04 (where available), 
ID from B97 or L02 (when applicable), RA, Dec, 
Radial velocity,
Velocity error, Major axis, Minor axis, Galactocentric radius,
Azimuthal angle, V, B$-$V, V$-$R, and Source (1=2dF, 2= WHT, 3= Keck, 
4= WIYN).\label{tab_globs}}
\tablewidth{508pt}
\tablehead{
RZ\_ID & Other\_ID & RA & Dec & Velocity & X & Y & R & $\theta$ & V & B$-$V & 
V$-$R & Source  \\
  &   & (2000) & (2000) & (km/s)  & ($\arcmin$) & ($\arcmin$) 
& ($\arcmin$) & ($\deg$) & (mag) & (mag) & (mag) & \\  
}
%\tableline
\startdata
5319 &  ...  &  12:39:00.69 & -11:35:18.0 & 1061 $\pm$  50 & -14.38 &  2.08 & 14.53 & 171.79 & 21.40 & 0.877 & 0.577 & 1\\
5289 &  ...  &  12:39:01.43 & -11:46:51.0 & 1092 $\pm$  58 & -14.19 & -9.47 & 17.06 & 213.73 & 20.11 & 0.636 & 0.467 & 1\\
4966 &  ...  &  12:39:11.62 & -11:34:00.6 & 1047 $\pm$  26 & -11.71 &  3.37 & 12.18 & 163.94 & 19.73 & 0.662 & 0.488 & 1\\
4884 &  ...  &  12:39:13.43 & -11:50:06.6 & 1238 $\pm$  57 & -11.25 & -12.73 & 16.99 & 228.52 & 20.60 & 0.649 & 0.473 & 1\\
4682 &  ...  &  12:39:18.51 & -11:30:57.5 & 1020 $\pm$  58 & -10.02 &  6.42 & 11.90 & 147.36 & 20.16 & 0.684 & 0.478 & 1\\
4679 &  ...  &  12:39:18.62 & -11:25:04.2 & 1063 $\pm$ 126 & -10.00 & 12.31 & 15.86 & 129.10 & 20.93 & 0.637 & 0.487 & 1\\
4303 &  ...  &  12:39:27.74 & -11:41:11.7 & 1070 $\pm$  31 & -7.76 & -3.81 &  8.64 & 206.18 & 19.91 & 0.802 & 0.516 & 1\\
4247 &  ...  &  12:39:29.37 & -11:29:33.3 & 1019 $\pm$  47 & -7.37 &  7.82 & 10.74 & 133.27 & 19.65 & 0.659 & 0.450 & 1\\
4206 &  ...  &  12:39:30.18 & -11:36:13.7 &  942 $\pm$  35 & -7.16 &  1.15 &  7.25 & 170.86 & 19.71 & 0.700 & 0.457 & 1\\
4158 &  ...  &  12:39:31.48 & -11:36:34.9 & 1285 $\pm$ 119 & -6.85 &  0.80 &  6.89 & 173.33 & 21.37 & 0.712 & 0.428 & 1\\
4108 &  ...  &  12:39:32.74 & -11:30:31.2 & 1206 $\pm$  64 & -6.54 &  6.86 &  9.48 & 133.64 & 20.18 & 0.735 & 0.509 & 1\\
3987 &  ...  &  12:39:35.63 & -11:38:30.1 & 1338 $\pm$  33 & -5.83 & -1.12 &  5.93 & 190.88 & 19.51 & 0.664 & 0.485 & 1\\
3974 &  ...  &  12:39:35.81 & -11:47:45.7 & 1031 $\pm$  72 & -5.78 & -10.38 & 11.88 & 240.89 & 20.61 & 0.683 & 0.472 & 1\\
3965 &  ...  &  12:39:36.08 & -11:34:08.6 & 1001 $\pm$  54 & -5.72 &  3.24 &  6.57 & 150.48 & 20.65 & 0.633 & 0.445 & 1\\
3892 &  ...  &  12:39:37.77 & -11:40:30.7 & 1136 $\pm$  32 & -5.30 & -3.13 &  6.15 & 210.56 & 19.57 & 0.908 & 0.580 & 1\\
3873 &  ...  &  12:39:38.14 & -11:38:38.5 &  840 $\pm$  58 & -5.21 & -1.26 &  5.36 & 193.58 & 20.16 & 0.691 & 0.463 & 1\\
3835 &  ...  &  12:39:38.91 & -11:35:12.2 & 1232 $\pm$  40 & -5.03 &  2.18 &  5.48 & 156.56 & 20.47 & 0.655 & 0.468 & 1\\
3728 &  1-20 &  12:39:41.19 & -11:36:08.3 & 1186 $\pm$  51 & -4.48 &  1.23 &  4.65 & 164.65 & 20.69 & 0.683 & 0.444 & 2\\
3680 &  ...  &  12:39:41.98 & -11:38:27.1 & 1099 $\pm$  25 & -4.27 & -1.07 &  4.40 & 194.05 & 19.47 & 0.705 & 0.484 & 1\\
3676 &  2-30 &  12:39:42.03 & -11:40:59.1 & 1104 $\pm$ 104 & -4.28 & -3.62 &  5.60 & 220.20 & 21.29 & 0.965 & 0.582 & 2\\
3667 &  1-39 &  12:39:42.14 & -11:35:23.4 &  832 $\pm$  61 & -4.25 &  1.98 &  4.69 & 155.02 & 21.31 & 0.831 & 0.528 & 2\\
3638 &  ...  &  12:39:42.75 & -11:42:48.0 &  987 $\pm$  45 & -4.08 & -5.42 &  6.78 & 232.98 & 20.38 & 0.666 & 0.475 & 1\\
3575 &  1-25 &  12:39:44.08 & -11:37:29.7 & 1231 $\pm$  26 & -3.78 & -0.12 &  3.78 & 181.90 & 20.75 & 0.842 & 0.580 & 2\\
3506 &  ...  &  12:39:45.62 & -11:39:33.3 &  946 $\pm$  28 & -3.38 & -2.17 &  4.02 & 212.71 & 19.47 & 0.960 & 0.623 & 1\\
3461 &  ...  &  12:39:46.61 & -11:36:50.6 & 1591 $\pm$  55 & -3.14 &  0.54 &  3.19 & 170.25 & 19.68 & 1.028 & 0.579 & 4\\
3394 &  ...  &  12:39:48.09 & -11:42:08.4 & 1330 $\pm$  31 & -2.78 & -4.76 &  5.51 & 239.72 & 19.96 & 0.724 & 0.470 & 1\\
3357 &  2-12 &  12:39:49.34 & -11:34:34.7 &  857 $\pm$  42 & -2.49 &  2.79 &  3.74 & 131.69 & 20.73 & 0.661 & 0.457 & 2\\
3320 &  ...  &  12:39:50.13 & -11:38:28.9 & 1259 $\pm$ 113 & -2.28 & -1.10 &  2.53 & 205.76 & 19.54 & 0.906 & 0.554 & 4\\
3284 &  1-32 &  12:39:50.83 & -11:38:54.4 & 1025 $\pm$ 203 & -2.12 & -1.54 &  2.62 & 216.00 & 21.11 & 0.808 & 0.484 & 2\\
3246 &  ...  &  12:39:51.41 & -11:44:06.4 & 1095 $\pm$  49 & -1.96 & -6.72 &  7.00 & 253.72 & 20.60 & 0.625 & 0.434 & 1\\
3243 &  2-31 &  12:39:51.52 & -11:38:16.1 &  939 $\pm$  21 & -1.95 & -0.90 &  2.15 & 204.69 & 21.33 & 0.997 & 0.628 & 2\\
3222 &  ...  &  12:39:51.89 & -11:36:27.1 & 1162 $\pm$  30 & -1.85 &  0.93 &  2.07 & 153.31 & 19.89 & 0.762 & 0.516 & 1\\
3191 &   2-4 &  12:39:52.37 & -11:35:40.4 &  968 $\pm$  78 & -1.75 &  1.69 &  2.43 & 135.85 & 20.41 & 0.877 & 0.564 & 2\\
3173 &  1-35 &  12:39:52.87 & -11:39:10.9 & 1300 $\pm$  40 & -1.62 & -1.81 &  2.43 & 228.19 & 21.00 & 0.661 & 0.483 & 2\\
 ... &  2-16 &  12:39:52.88 & -11:36:17.8 &  875 $\pm$  22 & -1.62 &  1.07 &  1.94 & 146.50 & 20.77 & 0.415 & 0.303 & 2\\
 ... &   2-7 &  12:39:54.07 & -11:36:04.9 &  976 $\pm$  24 & -1.33 &  1.29 &  1.85 & 135.93 & 20.52 & 0.859 & 0.566 & 2\\
3101 &  1-28 &  12:39:54.77 & -11:40:36.0 &  932 $\pm$  25 & -1.16 & -3.23 &  3.43 & 250.29 & 20.97 & 0.948 & 0.593 & 2\\
3093 &  ...  &  12:39:55.07 & -11:39:28.6 &  991 $\pm$  26 & -1.07 & -2.09 &  2.35 & 242.96 & 20.28 & 0.702 & 0.478 & 1\\
3075 &  ...  &  12:39:55.44 & -11:33:24.2 & 1148 $\pm$  31 & -0.98 &  3.98 &  4.10 & 103.84 & 19.81 & 0.717 & 0.471 & 1\\
3064 &  ...  &  12:39:55.73 & -11:39:32.3 &  972 $\pm$ 102 & -0.91 & -2.16 &  2.34 & 247.21 & 20.30 & 0.912 & 0.557 & 4\\
3061 &  ...  &  12:39:55.84 & -11:34:48.4 & 1007 $\pm$  30 & -0.88 &  2.58 &  2.72 & 108.95 & 19.46 & 0.997 & 0.611 & 1\\
 ... &  1-19 &  12:39:56.45 & -11:38:26.0 &  573 $\pm$  21 & -0.75 & -1.07 &  1.30 & 234.97 & 20.71 & 0.786 & 0.544 & 2\\
2985 &  ...  &  12:39:57.62 & -11:31:25.0 & 1277 $\pm$  27 & -0.45 &  5.97 &  5.98 &  94.28 & 19.28 & 0.916 & 0.599 & 1\\
 ... & C-136 &  12:39:58.03 & -11:37:59.6 & 1202 $\pm$  17 & -0.36 & -0.58 &  0.68 & 238.24 & 21.79 & 0.816 & 0.547 & 3\\
2916 &  ...  &  12:39:58.95 & -11:35:22.6 & 1074 $\pm$  24 & -0.12 &  2.01 &  2.01 &  93.41 & 19.53 & 0.735 & 0.524 & 1\\
 ... & C-132 &  12:39:59.28 & -11:38:21.5 &  679 $\pm$  15 & -0.05 & -0.94 &  0.94 & 266.73 & 20.99 & 0.871 & 0.542 & 3\\
 ... &   2-8 &  12:39:59.31 & -11:38:27.9 &  979 $\pm$  14 & -0.05 & -1.10 &  1.10 & 267.61 & 20.39 & 0.918 & 0.584 & 2\\
 ... & C-137 &  12:39:59.39 & -11:38:46.0 &  919 $\pm$  30 & -0.02 & -1.35 &  1.35 & 268.98 & 21.06 & 0.926 & 0.590 & 3\\
2877 &  2-13 &  12:39:59.82 & -11:34:54.6 &  616 $\pm$  34 &  0.08 &  2.46 &  2.46 &  88.17 & 20.89 & 0.592 & 0.487 & 2\\
2863 &  ...  &  12:40:00.16 & -11:32:27.7 & 1300 $\pm$  29 &  0.17 &  4.92 &  4.92 &  87.99 & 19.36 & 0.835 & 0.555 & 1\\
 ... & C-116 &  12:40:00.29 & -11:37:54.8 & 1022 $\pm$   8 &  0.19 & -0.50 &  0.54 & 291.15 & 20.12 & 0.716 & 0.501 & 3\\
2832 &  ...  &  12:40:00.59 & -11:44:56.0 & 1117 $\pm$  26 &  0.28 & -7.55 &  7.55 & 272.16 & 19.77 & 0.916 & 0.599 & 1\\
2828 &   2-6 &  12:40:00.70 & -11:35:19.6 & 1283 $\pm$  45 &  0.29 &  2.04 &  2.07 &  81.80 & 20.55 & 0.730 & 0.497 & 2\\
 ... & C-032 &  12:40:01.12 & -11:36:55.6 & 1055 $\pm$  24 &  0.40 &  0.48 &  0.62 &  50.53 & 20.60 & 0.747 & 0.524 & 3\\
2777 &  2-14 &  12:40:01.57 & -11:39:03.8 & 1045 $\pm$   7 &  0.51 & -1.70 &  1.77 & 286.72 & 20.95 & 0.869 & 0.557 & 2\\
 ... & C-051 &  12:40:01.58 & -11:36:16.9 & 1015 $\pm$  18 &  0.51 &  1.13 &  1.23 &  65.69 & 21.30 & 0.820 & 0.587 & 3\\
2748 &  ...  &  12:40:02.20 & -11:30:26.8 & 1150 $\pm$  36 &  0.67 &  6.94 &  6.97 &  84.45 & 19.87 & 0.740 & 0.479 & 1\\
2743 & H2-22 &  12:40:02.26 & -11:33:14.1 & 1274 $\pm$  17 &  0.65 &  4.11 &  4.16 &  81.07 & 18.65 & 0.566 & 0.336 & 3\\
 ... & C-042 &  12:40:02.57 & -11:37:08.9 & 1225 $\pm$  26 &  0.75 &  0.26 &  0.79 &  19.25 & 20.93 & 0.710 & 0.516 & 3\\
 ... & C-064 &  12:40:02.65 & -11:36:29.1 &  865 $\pm$  16 &  0.77 &  0.92 &  1.20 &  50.13 & 21.24 & 0.673 & 0.520 & 3\\
 ... & C-059 &  12:40:02.73 & -11:36:42.4 & 1034 $\pm$  19 &  0.79 &  0.70 &  1.06 &  41.67 & 21.32 & 0.933 & 0.566 & 3\\
 ... &  2-15 &  12:40:03.02 & -11:38:45.9 &  828 $\pm$  71 &  0.86 & -1.40 &  1.64 & 301.69 & 20.90 & 0.750 & 0.509 & 2\\
 ... & C-068 &  12:40:03.21 & -11:36:04.4 &  849 $\pm$  11 &  0.90 &  1.33 &  1.61 &  55.80 & 20.36 & 0.836 & 0.549 & 3\\
 ... & C-076 &  12:40:03.47 & -11:35:57.3 & 1563 $\pm$  77 &  0.97 &  1.45 &  1.74 &  56.22 & 21.51 & 0.607 & 0.498 & 3\\
2666 &  ...  &  12:40:03.66 & -11:34:50.1 & 1064 $\pm$ 137 &  1.04 &  2.55 &  2.75 &  67.86 & 19.76 & 0.858 & 0.547 & 4\\
2637 &  1-26 &  12:40:04.45 & -11:40:18.5 & 1199 $\pm$  14 &  1.21 & -2.94 &  3.18 & 292.37 & 20.86 & 0.914 & 0.535 & 2\\
 ... &  1-11 &  12:40:04.96 & -11:36:21.4 & 1457 $\pm$  12 &  1.34 &  1.01 &  1.67 &  37.07 & 20.71 & 0.642 & 0.450 & 2\\
2576 & H2-09 &  12:40:05.65 & -11:32:29.5 &  611 $\pm$   9 &  1.50 &  4.85 &  5.08 &  72.86 & 20.38 & 0.624 & 0.469 & 3\\
2551 &   2-2 &  12:40:06.31 & -11:36:47.2 &  808 $\pm$  39 &  1.67 &  0.58 &  1.77 &  19.29 & 20.40 & 1.040 & 0.568 & 2\\
2544 &  2-24 &  12:40:06.40 & -11:40:13.3 & 1505 $\pm$ 119 &  1.69 & -2.85 &  3.31 & 300.57 & 21.10 & 0.814 & 0.526 & 2\\
2512 &  ...  &  12:40:07.08 & -11:36:04.7 & 1189 $\pm$  27 &  1.87 &  1.30 &  2.28 &  34.90 & 19.67 & 0.920 & 0.579 & 1\\
2491 &   2-5 &  12:40:07.50 & -11:35:51.5 & 1220 $\pm$  89 &  1.96 &  1.51 &  2.47 &  37.67 & 20.49 & 0.751 & 0.496 & 2\\
2475 &  ...  &  12:40:07.68 & -11:46:04.8 & 1095 $\pm$  27 &  2.02 & -8.70 &  8.93 & 283.07 & 20.02 & 0.729 & 0.457 & 1\\
2449 &  ...  &  12:40:08.09 & -11:33:11.1 &  939 $\pm$  29 &  2.12 &  4.20 &  4.70 &  63.24 & 19.69 & 0.756 & 0.526 & 1\\
 ... &  2-11 &  12:40:08.51 & -11:33:57.3 & 1411 $\pm$  56 &  2.20 &  3.41 &  4.06 &  57.15 & 20.89 & 0.513 & 0.473 & 2\\
2421 &  ...  &  12:40:08.65 & -11:39:30.8 & 1270 $\pm$  36 &  2.26 & -2.13 &  3.10 & 316.66 & 19.85 & 0.917 & 0.572 & 4\\
2363 &  ...  &  12:40:09.73 & -11:36:22.6 & 1355 $\pm$  30 &  2.52 &  1.01 &  2.71 &  21.81 & 19.45 & 0.689 & 0.480 & 1\\
2358 &   1-4 &  12:40:09.83 & -11:35:20.6 & 1152 $\pm$  40 &  2.53 &  2.03 &  3.24 &  38.68 & 20.51 & 0.810 & 0.554 & 2\\
2316 &  ...  &  12:40:10.52 & -11:45:20.4 & 1115 $\pm$  35 &  2.71 & -7.96 &  8.41 & 288.84 & 20.31 & 0.864 & 0.545 & 4\\
2307 &   1-2 &  12:40:10.67 & -11:36:45.3 &  776 $\pm$  11 &  2.73 &  0.61 &  2.80 &  12.64 & 20.25 & 0.928 & 0.542 & 2\\
2303 &  ...  &  12:40:10.77 & -11:40:00.6 & 1207 $\pm$  34 &  2.77 & -2.63 &  3.82 & 316.55 & 20.02 & 0.696 & 0.456 & 1\\
2289 &   1-5 &  12:40:10.98 & -11:35:54.7 &  755 $\pm$  61 &  2.81 &  1.46 &  3.17 &  27.39 & 20.51 & 0.577 & 0.422 & 2\\
2257 &  2-17 &  12:40:11.38 & -11:39:42.9 & 1275 $\pm$ 100 &  2.91 & -2.35 &  3.74 & 321.06 & 20.82 & 0.698 & 0.473 & 2\\
2242 &   1-3 &  12:40:11.64 & -11:38:14.4 & 1369 $\pm$ 129 &  2.97 & -0.87 &  3.10 & 343.65 & 20.35 & 0.605 & 0.440 & 2\\
2237 &  1-15 &  12:40:11.74 & -11:41:56.5 & 1035 $\pm$  24 &  3.00 & -4.57 &  5.47 & 303.23 & 20.54 & 0.698 & 0.482 & 2\\
2200 &   2-3 &  12:40:12.45 & -11:38:12.1 & 1524 $\pm$  14 &  3.17 & -0.83 &  3.28 & 345.28 & 20.44 & 0.668 & 0.473 & 2\\
2203 &  ...  &  12:40:12.46 & -11:25:33.3 & 1174 $\pm$  32 &  3.19 & 11.83 & 12.25 &  74.92 & 19.55 & 0.644 & 0.479 & 1\\
2148 &  ...  &  12:40:13.48 & -11:35:51.3 & 1136 $\pm$  35 &  3.44 &  1.53 &  3.76 &  24.01 & 20.79 & 0.713 & 0.463 & 1\\
 ... &  1-29 &  12:40:13.59 & -11:39:12.0 &  853 $\pm$  31 &  3.45 & -1.83 &  3.91 & 332.02 & 20.97 & 1.079 & 0.672 & 2\\
2129 &  ...  &  12:40:13.79 & -11:37:02.4 & 1167 $\pm$  62 &  3.51 &  0.34 &  3.53 &   5.60 & 19.91 & 0.751 & 0.500 & 1\\
2109 &  ...  &  12:40:14.11 & -11:38:38.4 & 1350 $\pm$  41 &  3.59 & -1.26 &  3.81 & 340.74 & 20.50 & 0.821 & 0.539 & 1\\
2086 &  ...  &  12:40:14.64 & -11:31:55.5 & 1063 $\pm$  23 &  3.72 &  5.46 &  6.61 &  55.73 & 19.48 & 0.718 & 0.505 & 1\\
2038 &  ...  &  12:40:15.31 & -11:40:12.7 & 1241 $\pm$  27 &  3.89 & -2.83 &  4.81 & 323.97 & 19.33 & 0.753 & 0.487 & 1\\
2032 &  1-16 &  12:40:15.41 & -11:37:34.5 & 1256 $\pm$  18 &  3.90 & -0.21 &  3.90 & 356.93 & 20.62 & 0.874 & 0.620 & 2\\
2021 &  ...  &  12:40:15.58 & -11:36:21.0 & 1253 $\pm$  29 &  3.95 &  1.03 &  4.09 &  14.67 & 20.19 & 0.673 & 0.460 & 1\\
1771 &  ...  &  12:40:21.24 & -11:36:31.9 & 1181 $\pm$  35 &  5.34 &  0.85 &  5.41 &   9.07 & 20.82 & 0.751 & 0.507 & 1\\
1638 &  ...  &  12:40:24.74 & -11:39:50.7 & 1072 $\pm$  51 &  6.19 & -2.46 &  6.66 & 338.33 & 19.56 & 0.942 & 0.608 & 1\\
1430 &  ...  &  12:40:29.82 & -11:33:40.0 & 1171 $\pm$  43 &  7.44 &  3.72 &  8.32 &  26.55 & 20.19 & 0.687 & 0.474 & 1\\
1428 &  ...  &  12:40:29.82 & -11:49:00.2 & 1056 $\pm$  42 &  7.44 & -11.62 & 13.79 & 302.62 & 19.67 & 0.626 & 0.464 & 1\\
1255 &  ...  &  12:40:34.76 & -11:38:10.2 &  884 $\pm$  38 &  8.65 & -0.79 &  8.68 & 354.80 & 19.81 & 0.732 & 0.508 & 1\\
1233 &  ...  &  12:40:35.37 & -11:41:36.6 & 1230 $\pm$  52 &  8.80 & -4.23 &  9.76 & 334.34 & 21.41 & 0.918 & 0.585 & 1\\
1164 &  ...  &  12:40:36.77 & -11:30:25.8 &  773 $\pm$ 157 &  9.14 &  6.95 & 11.48 &  37.25 & 20.46 & 0.559 & 0.426 & 1\\
1154 &  ...  &  12:40:37.04 & -11:29:27.0 &  774 $\pm$  28 &  9.21 &  7.93 & 12.15 &  40.74 & 19.44 & 0.664 & 0.511 & 1\\
1150 &  ...  &  12:40:37.12 & -11:46:01.2 & 1046 $\pm$  23 &  9.22 & -8.64 & 12.63 & 316.88 & 19.29 & 0.740 & 0.443 & 1\\
1070 &  ...  &  12:40:39.00 & -11:55:01.9 & 1005 $\pm$  31 &  9.68 & -17.65 & 20.13 & 298.74 & 19.68 & 0.729 & 0.484 & 1\\
1045 &  ...  &  12:40:39.82 & -11:36:55.1 & 1025 $\pm$  26 &  9.89 &  0.46 &  9.90 &   2.68 & 19.67 & 0.759 & 0.483 & 1\\
 997 &  ...  &  12:40:41.23 & -11:34:33.0 &  818 $\pm$  22 & 10.23 &  2.83 & 10.62 &  15.48 & 19.18 & 0.767 & 0.538 & 1\\
 682 &  ...  &  12:40:52.22 & -11:50:52.7 &  681 $\pm$  54 & 12.92 & -13.49 & 18.68 & 313.75 & 21.01 & 0.694 & 0.440 & 1\\

\enddata
\end{deluxetable}

\begin{deluxetable}{ccccccc}
\tabletypesize{\scriptsize}
%\rotate
\tablewidth{300pt}
\tablecaption{Foreground Stars from 2dF and WIYN/Hydra
  Spectroscopy.
Successive columns give: ID from RZ04 (where available), RA, Dec,
Radial velocity and error,V, B$-$V, and V$-$R.  One object
(RZ\# 2458) was observed with WIYN/Hydra; all others are from the 2dF
data. 
\label{tab_nonGCs}}
%\tableline \tableline
\tablehead{
RZ\_ID & RA & Dec & Velocity & V & B$-$V & V$-$R \\
       & (2000) & (2000) & (km/s)  & (mag) & (mag) & (mag) \\  
}
%\tableline
\startdata
5635 &  12:38:50.85 & -11:42:09.2 &   92 $\pm$  33 & 20.87 & 0.735 & 0.462 \\
5605 &  12:38:51.48 & -11:48:53.4 &  177 $\pm$  33 & 19.59 & 0.552 & 0.402 \\
5574 &  12:38:52.88 & -11:39:19.9 &  117 $\pm$  45 & 20.98 & 0.912 & 0.548 \\
5516 &  12:38:54.50 & -11:49:22.3 &  148 $\pm$  26 & 19.71 & 0.887 & 0.610 \\
5489 &  12:38:55.37 & -11:27:26.8 &  303 $\pm$  56 & 20.38 & 0.651 & 0.392 \\
5423 &  12:38:57.19 & -11:41:58.9 &  139 $\pm$  67 & 20.60 & 0.611 & 0.425 \\
5390 &  12:38:58.04 & -11:51:45.3 &  270 $\pm$  23 & 19.25 & 0.613 & 0.428 \\
5315 &  12:39:00.88 & -11:28:42.9 &   10 $\pm$  30 & 19.26 & 0.984 & 0.647 \\
5275 &  12:39:02.17 & -11:21:02.0 &  434 $\pm$  94 & 20.00 & 0.502 & 0.396 \\
5192 &  12:39:04.61 & -11:23:24.3 &  456 $\pm$  62 & 20.58 & 0.551 & 0.379 \\
5184 &  12:39:04.64 & -11:50:53.3 &  166 $\pm$  28 & 19.89 & 0.904 & 0.578 \\
5097 &  12:39:07.56 & -11:43:04.1 &   94 $\pm$  29 & 19.33 & 0.638 & 0.418 \\
5023 &  12:39:09.54 & -11:53:04.0 &   68 $\pm$  43 & 19.64 & 0.529 & 0.350 \\
5024 &  12:39:09.64 & -11:27:35.9 &   99 $\pm$  33 & 19.41 & 1.003 & 0.688 \\
5006 &  12:39:10.18 & -11:46:22.0 &  -13 $\pm$  24 & 19.37 & 1.041 & 0.688 \\
4958 &  12:39:11.71 & -11:43:52.6 &   91 $\pm$  26 & 20.08 & 0.910 & 0.565 \\
4960 &  12:39:11.76 & -11:23:50.2 &   28 $\pm$  32 & 19.12 & 1.036 & 0.706 \\
4800 &  12:39:15.82 & -11:20:23.5 &  142 $\pm$  26 & 19.43 & 0.833 & 0.547 \\
4766 &  12:39:16.30 & -11:53:48.6 &  236 $\pm$  32 & 19.84 & 0.849 & 0.512 \\
4728 &  12:39:17.39 & -11:51:24.4 &   95 $\pm$  21 & 19.23 & 0.878 & 0.563 \\
4671 &  12:39:18.73 & -11:40:04.3 &  405 $\pm$  51 & 21.11 & 0.774 & 0.506 \\
4647 &  12:39:19.26 & -11:47:52.8 &  307 $\pm$  37 & 20.01 & 0.694 & 0.389 \\
4553 &  12:39:21.57 & -11:37:21.0 &  130 $\pm$  27 & 19.29 & 0.774 & 0.576 \\
4530 &  12:39:22.09 & -11:46:02.4 &  -20 $\pm$  30 & 19.97 & 0.832 & 0.514 \\
4404 &  12:39:25.45 & -11:45:21.6 &   47 $\pm$  49 & 21.03 & 1.020 & 0.632 \\
4318 &  12:39:27.50 & -11:38:05.3 &  -38 $\pm$  42 & 19.48 & 0.613 & 0.405 \\
4287 &  12:39:28.27 & -11:52:44.8 &  139 $\pm$  33 & 20.55 & 0.954 & 0.594 \\
4201 &  12:39:30.42 & -11:33:06.1 &   64 $\pm$  39 & 19.29 & 0.648 & 0.418 \\
4180 &  12:39:30.89 & -11:26:10.4 &  158 $\pm$  19 & 19.21 & 0.823 & 0.538 \\
4064 &  12:39:33.78 & -11:23:08.8 &  236 $\pm$  43 & 19.39 & 0.514 & 0.383 \\
4037 &  12:39:34.48 & -11:36:54.5 &   74 $\pm$  42 & 19.93 & 0.628 & 0.408 \\
3994 &  12:39:35.39 & -11:54:32.1 &   53 $\pm$  23 & 19.76 & 0.878 & 0.550 \\
3948 &  12:39:36.62 & -11:19:16.1 &   66 $\pm$  26 & 19.18 & 1.013 & 0.738 \\
3880 &  12:39:38.10 & -11:20:20.5 &  -24 $\pm$  31 & 19.83 & 0.831 & 0.566 \\
3830 &  12:39:39.05 & -11:33:16.6 &  142 $\pm$  46 & 19.91 & 0.607 & 0.393 \\
3787 &  12:39:39.88 & -11:51:38.9 &  232 $\pm$  41 & 20.86 & 0.909 & 0.547 \\
3684 &  12:39:41.90 & -11:52:27.9 &  297 $\pm$  40 & 19.59 & 0.517 & 0.359 \\
3669 &  12:39:42.16 & -11:21:09.4 &  143 $\pm$  30 & 20.21 & 0.988 & 0.670 \\
3650 &  12:39:42.46 & -11:49:04.8 &  127 $\pm$  28 & 20.09 & 1.070 & 0.664 \\
3543 &  12:39:44.58 & -11:38:26.8 &   49 $\pm$  28 & 19.66 & 0.968 & 0.609 \\
3478 &  12:39:46.28 & -11:20:20.8 &  -37 $\pm$  31 & 19.88 & 1.050 & 0.693 \\
3448 &  12:39:46.85 & -11:46:13.8 &   34 $\pm$  27 & 19.15 & 0.984 & 0.693 \\
3387 &  12:39:48.31 & -11:44:34.3 &  122 $\pm$  55 & 19.79 & 0.728 & 0.448 \\
3329 &  12:39:49.92 & -11:46:57.3 &  156 $\pm$  24 & 19.52 & 0.649 & 0.377 \\
3264 &  12:39:51.25 & -11:26:34.1 &  203 $\pm$  28 & 20.08 & 0.637 & 0.404 \\
3223 &  12:39:51.82 & -11:50:38.1 &   33 $\pm$  29 & 20.05 & 0.642 & 0.396 \\
3215 &  12:39:52.01 & -11:22:57.3 &  331 $\pm$  23 & 19.45 & 0.755 & 0.527 \\
3140 &  12:39:53.78 & -11:49:36.9 &  351 $\pm$  46 & 20.11 & 0.574 & 0.337 \\
2931 &  12:39:58.63 & -11:42:32.9 &   -7 $\pm$  38 & 19.61 & 0.586 & 0.432 \\
2855 &  12:40:00.13 & -11:55:04.4 &  211 $\pm$  82 & 20.38 & 0.565 & 0.352 \\
2849 &  12:40:00.31 & -11:48:29.7 &  369 $\pm$  49 & 21.07 & 0.804 & 0.473 \\
2803 &  12:40:01.22 & -11:31:17.7 &  178 $\pm$  47 & 19.95 & 0.506 & 0.344 \\
2775 &  12:40:01.56 & -11:52:11.0 &  370 $\pm$  27 & 19.59 & 0.673 & 0.398 \\
2582 &  12:40:05.58 & -11:26:44.4 &    0 $\pm$  24 & 19.46 & 0.707 & 0.455 \\
2483 &  12:40:07.61 & -11:39:50.9 &  205 $\pm$  45 & 19.78 & 0.532 & 0.334 \\
2458 &  12:40:07.89 & -11:50:45.9 &  -20 $\pm$ 122 & 19.53 & 0.974 & 0.602 \\
2349 &  12:40:10.01 & -11:20:23.2 &   67 $\pm$  31 & 19.41 & 1.079 & 0.707 \\
2247 &  12:40:11.48 & -11:52:47.4 &  155 $\pm$  30 & 19.17 & 0.511 & 0.345 \\
2221 &  12:40:11.94 & -11:44:52.3 &   94 $\pm$  28 & 19.82 & 0.893 & 0.546 \\
2098 &  12:40:14.32 & -11:23:40.8 &    4 $\pm$  28 & 19.71 & 0.775 & 0.500 \\
1973 &  12:40:16.86 & -11:39:19.2 &   40 $\pm$  22 & 19.36 & 0.755 & 0.444 \\
1950 &  12:40:17.34 & -11:21:23.2 &  157 $\pm$  92 & 20.17 & 0.732 & 0.425 \\
1908 &  12:40:18.12 & -11:54:27.1 &  113 $\pm$  24 & 19.45 & 0.630 & 0.427 \\
1775 &  12:40:21.17 & -11:24:37.9 &  106 $\pm$  29 & 19.95 & 0.880 & 0.584 \\
1773 &  12:40:21.20 & -11:33:43.5 &  211 $\pm$  28 & 19.28 & 0.599 & 0.416 \\
1695 &  12:40:23.21 & -11:30:56.7 &   22 $\pm$  24 & 19.50 & 1.075 & 0.720 \\
1577 &  12:40:26.09 & -11:33:56.8 &  296 $\pm$  40 & 19.86 & 0.557 & 0.393 \\
1464 &  12:40:28.92 & -11:26:52.0 &  307 $\pm$  92 & 20.76 & 0.516 & 0.332 \\
1429 &  12:40:29.83 & -11:19:59.5 &  228 $\pm$  33 & 19.78 & 0.627 & 0.426 \\
1278 &  12:40:33.96 & -11:39:35.3 &  167 $\pm$  30 & 19.13 & 0.951 & 0.611 \\
1256 &  12:40:34.74 & -11:50:18.6 &  102 $\pm$  46 & 20.28 & 0.971 & 0.629 \\
1167 &  12:40:36.74 & -11:43:06.8 &   44 $\pm$  45 & 19.38 & 0.562 & 0.380 \\
1149 &  12:40:37.14 & -11:23:44.8 &   96 $\pm$  45 & 21.19 & 0.932 & 0.601 \\
1060 &  12:40:39.31 & -11:52:08.0 &  -15 $\pm$  36 & 19.83 & 1.070 & 0.705 \\
1059 &  12:40:39.33 & -11:41:16.8 &   62 $\pm$  30 & 19.29 & 1.073 & 0.691 \\
1030 &  12:40:40.23 & -11:19:23.8 &  178 $\pm$  47 & 20.29 & 0.634 & 0.429 \\
1005 &  12:40:41.07 & -11:54:11.9 &   63 $\pm$  28 & 19.65 & 0.785 & 0.540 \\
1003 &  12:40:41.12 & -11:30:50.0 &   63 $\pm$  20 & 19.09 & 0.923 & 0.609 \\
 939 &  12:40:43.60 & -11:19:49.5 &    7 $\pm$  29 & 19.47 & 0.977 & 0.660 \\
 922 &  12:40:44.01 & -11:46:40.8 &  233 $\pm$  60 & 20.76 & 0.814 & 0.528 \\
 919 &  12:40:44.10 & -11:34:45.9 &  278 $\pm$  43 & 20.06 & 1.041 & 0.680 \\
 906 &  12:40:44.44 & -11:25:43.5 &  314 $\pm$  49 & 19.49 & 0.520 & 0.375 \\
 893 &  12:40:44.90 & -11:43:02.4 &   27 $\pm$  31 & 19.53 & 0.899 & 0.569 \\
 891 &  12:40:44.92 & -11:55:01.7 &   38 $\pm$  37 & 19.09 & 0.763 & 0.458 \\
 866 &  12:40:45.84 & -11:31:53.8 &   78 $\pm$  18 & 19.04 & 0.575 & 0.391 \\
 722 &  12:40:51.06 & -11:34:56.7 &   28 $\pm$  21 & 19.58 & 0.728 & 0.470 \\
 718 &  12:40:51.20 & -11:47:06.3 & -143 $\pm$  66 & 20.73 & 0.577 & 0.448 \\
 714 &  12:40:51.37 & -11:42:48.9 &  240 $\pm$  39 & 19.42 & 0.789 & 0.481 \\
 708 &  12:40:51.53 & -11:28:43.2 &  128 $\pm$  30 & 19.47 & 0.884 & 0.643 \\
 661 &  12:40:52.84 & -11:51:34.1 &  399 $\pm$  56 & 20.30 & 0.609 & 0.402 \\
 630 &  12:40:53.76 & -11:37:07.7 &  222 $\pm$  29 & 19.70 & 0.577 & 0.393 \\
 573 &  12:40:55.52 & -11:49:12.2 &   -7 $\pm$  21 & 19.25 & 0.869 & 0.560 \\
 572 &  12:40:55.57 & -11:33:32.7 &  -34 $\pm$  20 & 19.52 & 0.676 & 0.444 \\
 550 &  12:40:56.33 & -11:45:23.6 &  193 $\pm$  49 & 20.19 & 0.725 & 0.430 \\
 449 &  12:40:59.79 & -11:53:55.0 &  205 $\pm$  59 & 20.62 & 0.830 & 0.578 \\
 442 &  12:41:00.00 & -11:36:15.4 &    0 $\pm$  20 & 19.27 & 0.676 & 0.427 \\
 436 &  12:41:00.14 & -11:36:49.0 &  137 $\pm$  29 & 19.43 & 0.801 & 0.422 \\
 425 &  12:41:00.36 & -11:32:57.0 &   32 $\pm$  26 & 19.51 & 0.986 & 0.688 \\
 399 &  12:41:01.39 & -11:25:14.4 &  101 $\pm$  26 & 19.65 & 0.958 & 0.623 \\
 386 &  12:41:01.65 & -11:27:06.0 &  221 $\pm$  25 & 19.21 & 0.674 & 0.449 \\
 293 &  12:41:06.35 & -11:19:56.1 &  143 $\pm$  22 & 19.34 & 0.733 & 0.432 \\
 253 &  12:41:08.12 & -11:50:46.7 &  168 $\pm$  23 & 19.89 & 0.853 & 0.584 \\
 233 &  12:41:08.76 & -11:27:55.6 &  293 $\pm$  35 & 19.56 & 0.591 & 0.409 \\
 228 &  12:41:09.06 & -11:40:46.2 &  296 $\pm$  22 & 19.54 & 0.713 & 0.430 \\
 136 &  12:41:12.22 & -11:47:57.0 &   57 $\pm$  30 & 19.99 & 0.933 & 0.655 \\
  19 &  12:41:15.59 & -11:41:58.5 &  109 $\pm$  19 & 19.08 & 0.715 & 0.512 \\
  16 &  12:41:15.78 & -11:45:40.7 &   98 $\pm$  21 & 19.30 & 0.689 & 0.488 \\
   9 &  12:41:15.88 & -11:33:37.2 &   68 $\pm$  30 & 20.02 & 0.933 & 0.716 \\

\enddata
\end{deluxetable}

\clearpage

\section{Results}
\label{results}

\subsection{Rotation of the Globular Cluster System}

The wide field of our study allows us to test for the presence or
absence of rotation in the M104 GC system out to 
10\arcmin\ in radius ($\sim$ 30 kpc).
In the inner regions, we can compare the results
for the GC system to the rotation seen in the stars and gas, while at
large radius, our GC data provide a unique probe for rotation in
the outer halo of a bulge-dominated galaxy.

\begin{figure}
\epsscale{1.0}
\plotone{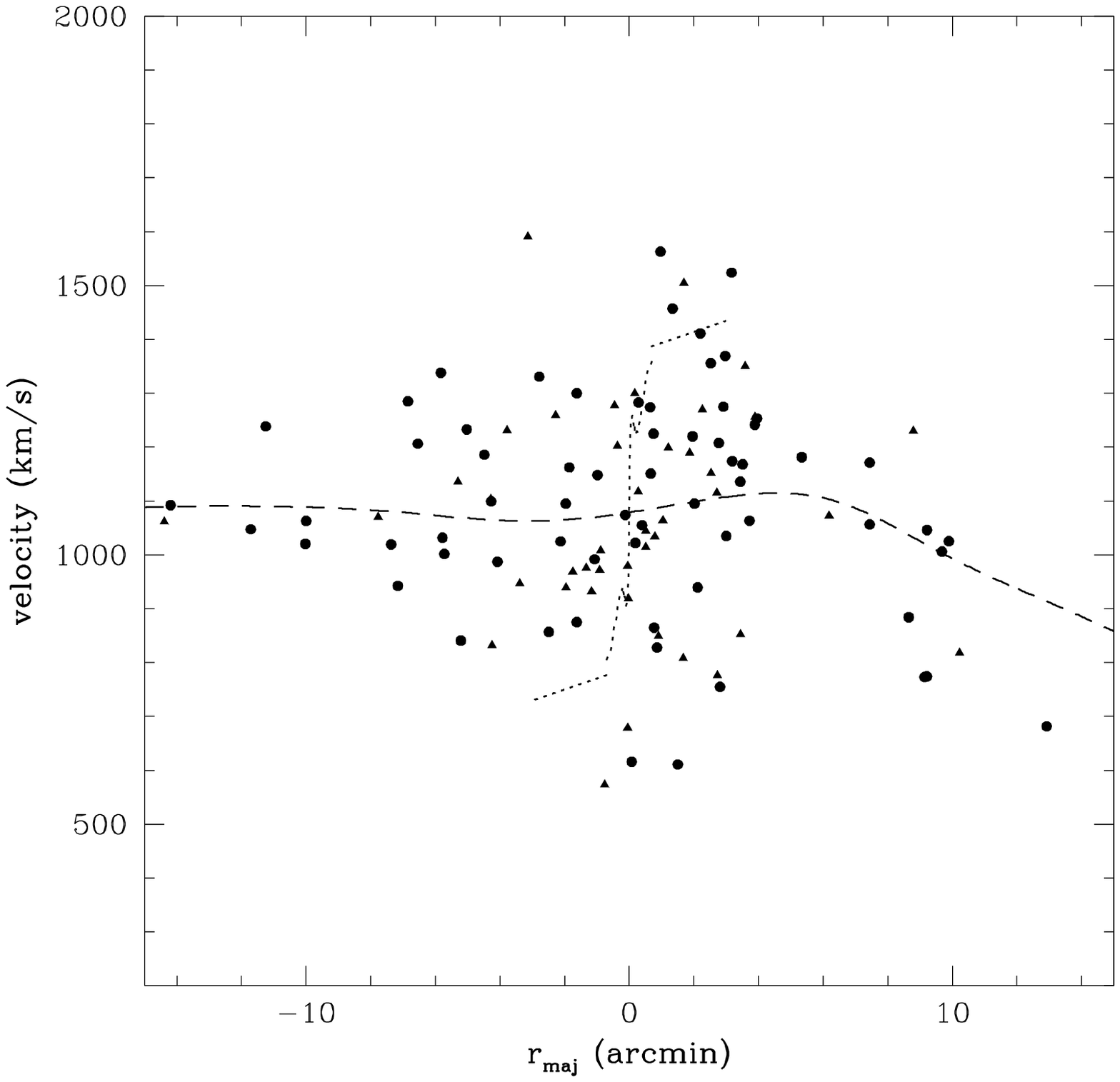}
\caption{Radial velocity vs major axis distance for M104 GCs. 
Circles are blue GCs (B$-$R $<$ 1.3), and triangles are red
GCs (B$-$R $>$ 1.3). 
The dashed line is a smoothed fit to the GC velocities, 
while the dotted line shows the stellar and gas rotation curve
from van der Marel et al. (1994) and Kormendy \& Westpfahl (1989).}
\label{fig:vel_major}
\end{figure}

Given the strong rotation in the gas and stars in the disk of
M104, it is a natural first step to search for rotation along
the major axis of the galaxy. In Figure \ref{fig:vel_major} we
plot radial velocity vs.\ major axis distance for the 108 GCs 
in our sample. It is immediately apparent from this figure that
the GC system has little rotation along the major axis, in contrast
to the large rotation seen in the stars and gas along the disk 
on the major axis. This rotation from major axis measurements 
of stars and gas is shown as the dotted line, where the inner
points to $r \simeq 0.7'$ are from the stellar absorption line
measurements of van der Marel et al. (1994) and the line beyond this
from the fit of Kormendy \& Westpfahl (1989) to a number of
observations of HII regions and HI gas, which have their farthest 
extent at $r \simeq 3'$. 

The smoothed velocity profile for the GC sample is shown in 
Figure \ref{fig:vel_major} as the dashed line, where the
velocities have been smoothed using a Gaussian kernel with a
width of $\sigma = 3'$. This confirms the strong visual
impression that there is no clear rotation in the GC sample out 
to large radii. Based on a smaller dataset
at $r < 5.5\arcmin\ $, B97 presented a 
tentative detection of rotation (at a confidence level of 92.5\%),
that is not found in our more extensive work here.
Figure \ref{fig:vel_major} also shows that 
the metal-rich and metal-poor GCs are similar in their weak 
or absent rotation. There is a hint of counter-rotation 
or asymmetry in the velocity profile at very large radius, but 
spectroscopic data for significantly more outer GCs will be 
required to test any such effect.  The dominant
conclusion from the GC data is that there is no significant
rotation over the large range of radii studied, and any rotation
that might be present is much smaller than the observed rotation
of stars and gas along the major axis. 

\begin{figure}
\epsscale{1.0}
%\epsangle{-90}
\plotone{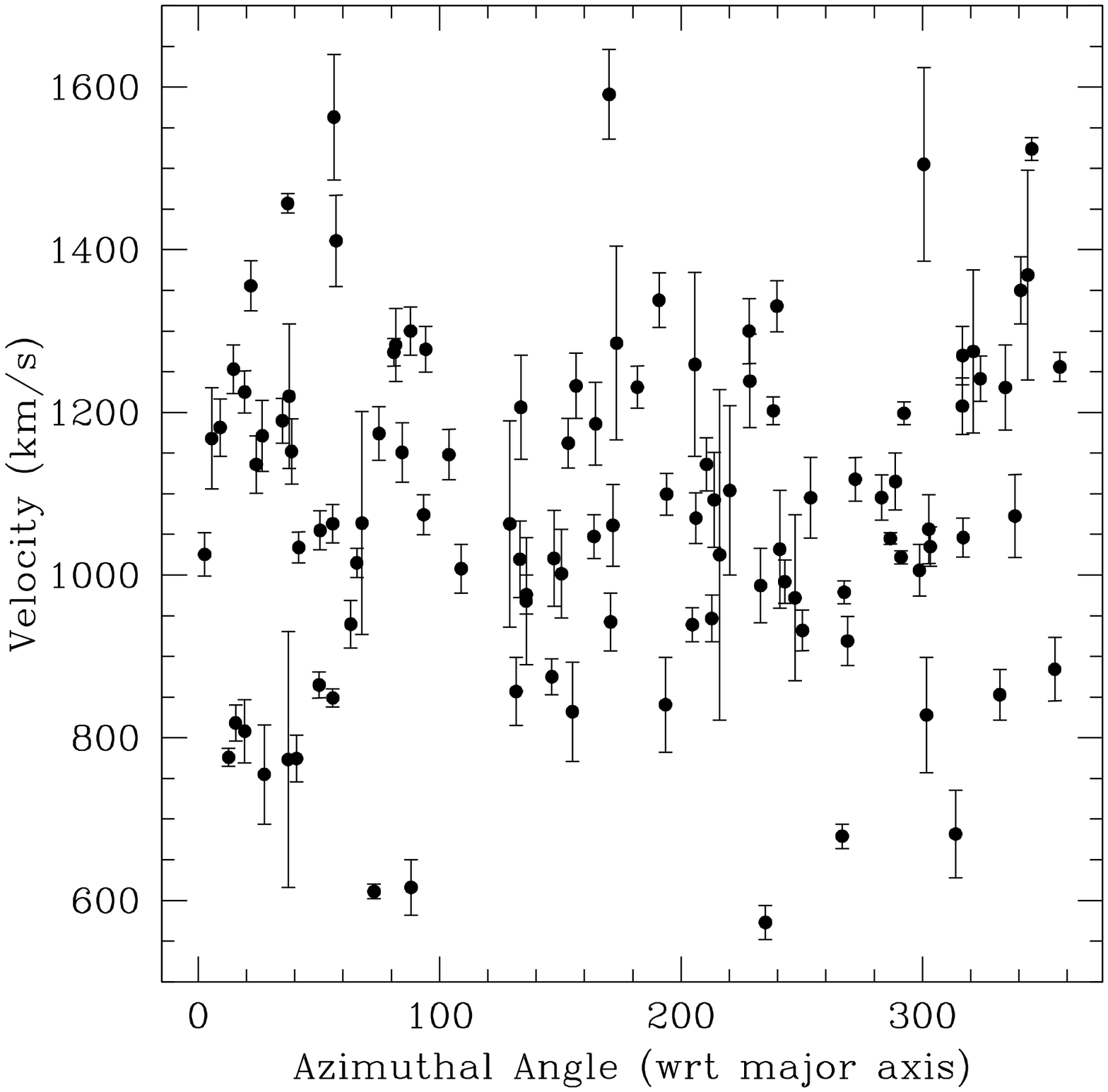}
\caption{Radial velocity vs azimuthal angle for M104 GCs.
There is no clear rotation seen in the M104 GC system.}
\label{fig:rotation_all}
\end{figure}

We can also extend our search for rotation to all possible
position angles.
Figure \ref{fig:rotation_all} plots radial velocity vs. azimuthal
angle for all 108 M104 GCs. Figure \ref{fig:rotation_all} shows 
that there is no obvious rotation about any position angle in 
the M104 GC system. To
quantify this result, we have performed non-linear least squares fits
to this equation:
V($\theta$) = V$_{rot}$sin($\theta$ - $\theta_0$) + V$_0$, 
where $V_{rot}$ is the rotation amplitude, $\theta$ is the
azimuthal angle ($\theta$ = 0 corresponds to the positive branch
of the major axis, and East on the sky), and $V_0$ is the systemic
velocity of M104. This corresponds to determining
the best-fitting flat rotation curve (see Zepf et al. 2000 for more
details).

We carried out this test for rotation about any axis in the total 
GC sample, and also in sub-samples split by colour and galactocentric 
radius. Our colour boundary is set at (B$-$R) = 1.3 (see Section \ref{id}),
while we divide by radius at 5\arcmin\ ($\sim$ 15 kpc). 
Columns 2 \& 3 of Table \ref{tab:rotation}
give the best-fit rotation velocity and position angle
returned by the least-squares code.
We have assessed the significance of these rotation velocities through
Monte-Carlo simulations which keep the GC azimuthal angles but randomize
the velocities (cf. Zepf et al. 2000); the significance level is defined
as the fraction of simulations with rotation velocity lower than the 
best-fit rotation velocity. As can be seen from Column 4 of 
Table \ref{tab:rotation},
there is no significant rotation seen in any of the GC sub-samples. We
have also used Monte Carlo simulations to set upper limits on the rotation
velocity. We do this by generating artificial samples with the same
position angle distributions and velocity dispersions as the data, with
a rotation curve of given amplitude imposed. The 95\% confidence
upper limit is defined as the rotation velocity for which only 5\%
of these simulations give a rotation velocity as small as that observed
(see Zepf et al. 2000); this upper limit is given in Column 5
of Table \ref{tab:rotation}. We then use these values, together with
the measured velocity dispersion of each sample (Column 6), to 
determine upper limits for (v/$\sigma$), which measures the importance
of rotation in a dynamical system. Our (v/$\sigma$) upper limits,
given in the 7th Column of Table \ref{tab:rotation}, range from 0.5 to 1.6.

The lack of rotation seen in the M104 GC system is surprising. Given the
edge-on alignment of M104's disk, it is difficult to argue that
there is rotation in the GC system but that we are viewing it face-on.
Cosmological simulations predict that early-type galaxies have
significant amounts of angular momentum (e.g. Vitvitska et al. 2002).
The stellar and gas
disk clearly has some angular momentum (Figure \ref{fig:vel_major}),
but the disk is only 10\% of the total $B$ light
(Burkhead 1986), and less by mass, and cannot account for the
expected overall angular momentum. There would seem to be two options:
that some luminous spheroid-dominated galaxies have less angular momentum
than expected from hierarchical merging; or that angular momentum has
been transferred to even larger radius by mergers during the formation
of M104 (e.g. Hernquist \& Bolte 1993). Given the long-standing and
consistent predictions for the angular momentum content of galaxies,
the latter option of angular momentum transport 
through merger-like processes would seem to be favoured.
The absence of evidence for
{\it recent} activity in M104 from either GC ages (L02; Hempel et al. 2006) 
or other measures
would suggest these processes happened in M104 a number of Gyr ago,
corresponding to redshifts greater than one or two.
Interestingly, the M104 bulge planetary nebulae (PNe) show spheroidal
rotation-- between a radius of 1$-$4\arcmin, the PNe rotation 
is $\sim$ 100 km/s near the equatorial plane,
and decreasing away from the plane (K. Freeman, unpublished data). 
The PNe rotation, coupled with the lack of rotation in the GC system, show
that these two populations are dynamically quite distinct, and presumably
formed in very different ways.
We plan to obtain more GC velocities at very large radii to further 
investigate these issues.

\begin{table}[h]
\caption{Rotation analysis for M104 GCs. Column 1 gives the sample
under consideration, Column 2 the least-squares fit for the rotation
velocity, Column 3 the position angle for this fit,
Column 4 the significance of this fit, Column 5 the
95\% upper limit on the rotation velocity, Column 6 the velocity
dispersion, and Column 7 the 95\% upper limit on (v/$\sigma$).}
\label{tab:rotation}
\begin{tabular}{lcccccc}
\hline
Sample & V$_{rot}$ & $\theta_0$ & Significance & V$^{max}_{rot}$ (95\%) &
$\sigma$ & (v/$\sigma$) (95\%) \\
 & (km/s) & (deg) & (\%) & (km/s) & (km/s) & \\
\hline
All clusters (N=108)   & 60  & -41 & 35 & 100 & 204 & 0.49 \\
(B$-$R $<$ 1.3) (N=66) & 120 & -38 & 36 & 175 & 203 & 0.86 \\
(B$-$R $>$ 1.3) (N=42) & 30  & 35 & 13 & 100 & 207 & 0.48 \\
R $<$ 5\arcmin\ (N=65) & 82   & 7 & 57 & 140 & 233 & 0.60 \\
R $>$ 5\arcmin\ (N=43) & 198   & 60 & 81 & 250 & 155 & 1.61 \\
\hline
\end{tabular}
\end{table}

\subsection{Globular Cluster Velocity Dispersion Profile}

\begin{figure}
\epsscale{1.0}
\plotone{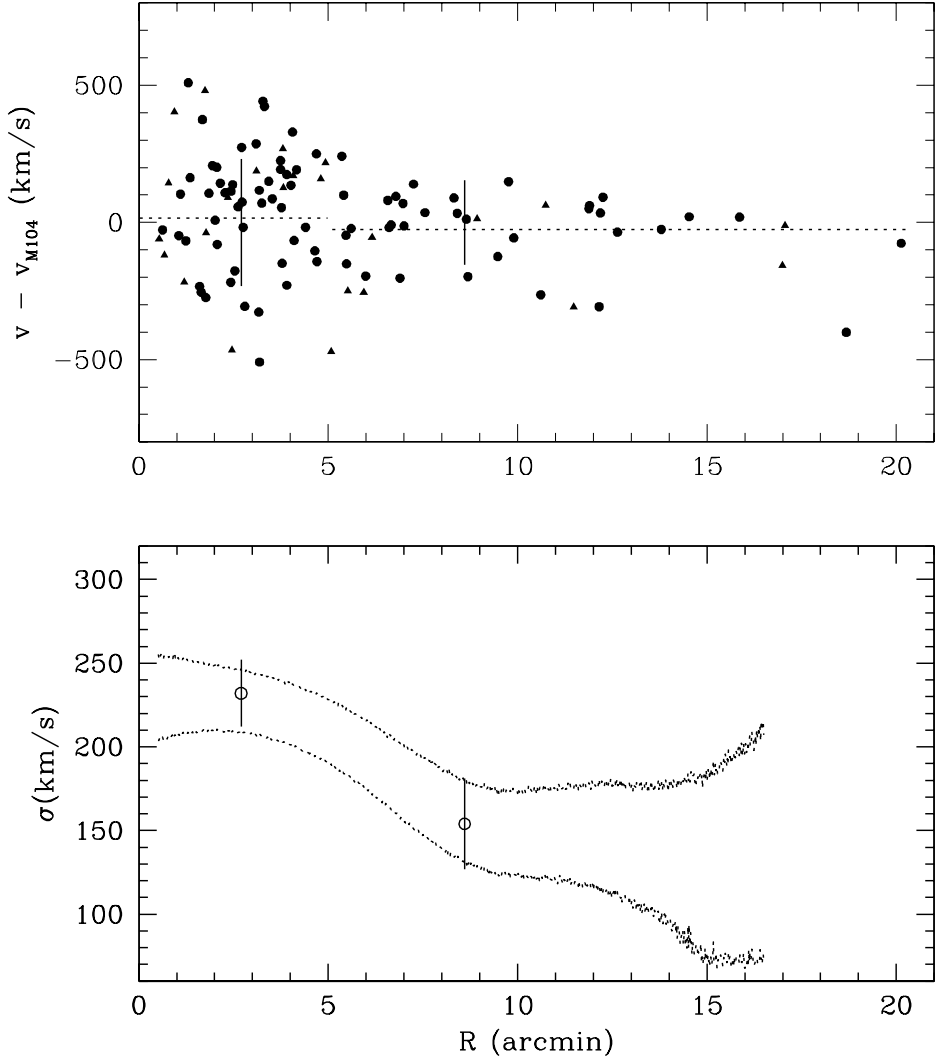}
\caption{{\bf Top:} V $-$ V$_{M104}$ vs. projected radius R 
for M104 GCs ($R$). Metal-rich GCs are plotted 
as triangles and metal-poor as circles, as in 
Figure \ref{fig:vel_major}. The 
dotted horizontal lines show the mean velocity for inner $(R < 5')$ 
and outer $(R > 5'$) samples, while the solid vertical lines indicate the
velocity dispersion for each of these subsamples. We adopt a systemic
velocity V$_{M104}$ = 1083 km/s, the mean velocity for all GCs.
The radial extent of this figure is about 60 kpc (12 $R_e$). 
{\bf Bottom:} Smoothed velocity dispersion profile for
M104 GCs. The upper and lower dotted lines represent the
$1\sigma$ limits for the velocity dispersion. The open circles
with error bars are the dispersions in the two bins mentioned
above, plotted at the median radius of each of the bins.
This plot shows the
declining velocity dispersion profile seen in our data.}
\label{fig:dispersion}
\end{figure}

The large radial extent of our GC sample also
allows us to probe the velocity dispersion profile of the M104
GC system from the inner region well out into the distant halo.
This dispersion profile can then be combined with the observed
spatial profile of the GC system and assumptions
about the GC orbits to derive the mass distribution of M104
over a large radial range.
To display the data, we plot in the top panel of 
Figure \ref{fig:dispersion} V$-$V$_{M104}$ against 
projected radius R for our 108 M104 GCs. Note that the mean
velocities for GCs with R $<$ 5\arcmin\ (65 GCs) and R $>$ 5\arcmin\ (43 GCs),
1099 $\pm$ 28 and 1058 $\pm$ 25 km/s respectively, agree with the
mean velocity for all GCs (1083 $\pm$ 20 km/s).
To enable any rotation
to be made clear, we have flipped the sign of the offset for GCs 
to the west of the galaxy center (those with negative $r_{maj}$ 
in Figure \ref{fig:vel_major}). 
Thus the overall consistency of the offset velocities 
with zero in Figure \ref{fig:dispersion} is further evidence for
the absence of rotation discussed earlier. The exception to this
may be for R $>$ $\sim$ 10\arcmin, where most of the points have
V$-$V$_{M104}$ $<$ 0 (effectively counter-rotating with respect to
the stars and gas in the inner galaxy). As mentioned 
in the previous section, more data will be
required to obtain a definitive result on this issue.

The top panel of Figure \ref{fig:dispersion} suggests that
the velocity dispersion of the M104 GCs declines with projected radius,
with the impression that it is fairly high inside of $\sim 5'$,
then declines fairly steeply, and possibly flattens at large radii.
We first quantify this by dividing our sample into an inner and
an outer bin, with a somewhat arbitrary division at R=5\arcmin.
The dispersions in
these two regions are $\sigma$(R$<$5\arcmin) = 233 $\pm$ 20 km/s,
and $\sigma$(R$>$5\arcmin) = 155 $\pm$ 28 km/s.
We can also make a continuous estimate of the velocity dispersion
with radius by smoothing the individual GC velocities.
In addition to avoiding issues of binning, this calculation
of the smoothed velocity dispersion profile allows the mass
distribution to be determined. The bottom panel of 
Figure \ref{fig:dispersion} shows the smoothed velocity
dispersion profile vs. projected radius $R$ for our
M104 dataset. The smoothing is done with a Gaussian kernel
with a width that slowly increases from $\sigma$ = 2\arcmin\ in the
center to $\sigma$ = 2.5\arcmin\ in the outer region, and the
uncertainties in the dispersion estimate are based on a 
bootstrap technique accounting for the kernel used (cf. Zepf et al. 2000).
The upper and lower dotted lines correspond respectively to the 
$1\sigma$ upper and lower limits for the velocity dispersion at that radius.
We have not corrected our velocity dispersions for measurement errors;
these corrections would change the velocity dispersion by 
less than 10\% everywhere, except perhaps at the very largest radii 
if the true dispersion is very low and the typical measurement 
uncertainties are at the upper end of published estimates.
The binned results discussed above
are plotted as open circles with error bars, where the points
are plotted at the $R$ corresponding to the median distance
of the GC sample in each bin from the galaxy center. 

The primary result from this analysis is that the dispersion is 
about 230 km/s
within the central few arcmin (recall that $R_e$ = 105\arcsec\ or 5 kpc),
and then begins to decline fairly steeply at around $5'$, and
then possibly flattens at larger radii, although our data are
too sparse at very large radii (e.g. $R \sim 15\arcmin$ or $\sim 45$ kpc)
to tell. A key use of this dispersion profile is for estimating
the mass of M104 through the Jeans equation, which we do in the
following section.

\subsection{The Mass Distribution of M104}
\label{dmhalo}

The velocity dispersion profile determined above can be combined
with the previously published spatial profile of the GC system
(RZ04) and the Jeans equation to determine the mass distribution
of M104. The large radial range of our GC velocity data
allows us to compare the mass estimate from the GCs with other
tracers out to a few arcminutes, and then to extend the
estimate of the mass distribution much farther into the halo
of M104. This calculation of the mass distribution from the 
Jeans equation is straightforward with the assumptions of
isotropy and spherical symmetry. Spherical symmetry is supported at
least outside the central regions by the round isophotes
of M104 (e.g.\ Burkhead 1986). Isotropy for the GC orbits in
ellipticals is supported by all cases in which an independent
constraint on the mass distribution from X-ray observations is
available (e.g. M49: Zepf et al. 2000, C\^ot\'e et al. 2003; 
M87: C\^ot\'e et al. 2001; M60: Bridges et al. 2006), although the 
metal-poor and metal-rich GCs in these galaxies 
often show differences from isotropy when considered separately.
Moreover, for our work on M104, we can test the mass distribution
with these assumptions directly against the mass inferred from the
rotation curve measured out to $r \le 3'$.

\begin{figure}
\plotone{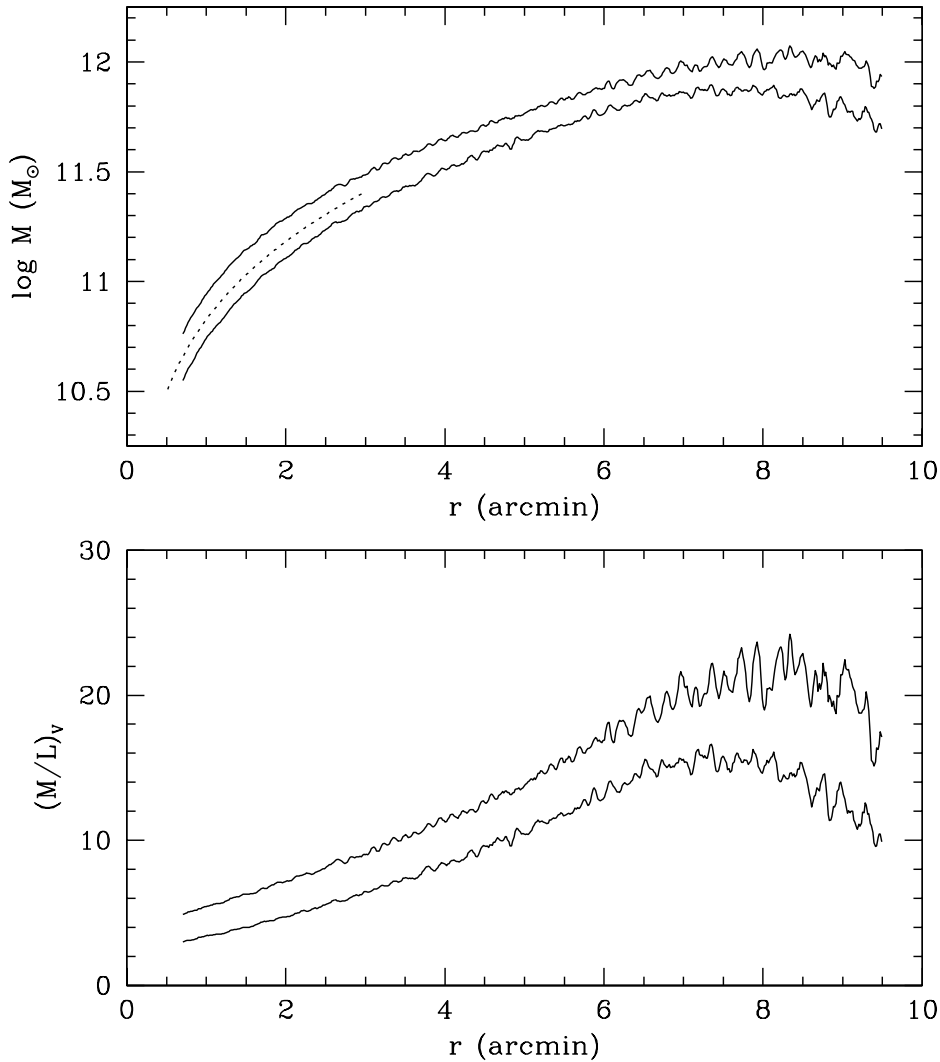}
\caption{
{\bf Top:} Mass vs deprojected radius r for M104. The solid lines give
the 1$\sigma$ bounds for the mass profile as determined from the
GCs, while the dotted line gives the mass profile determined from
the stellar rotation curve (Kormendy \& Westpfahl 1989). {\bf Bottom:} 
M/L$_V$ vs deprojected radius r for M104. The mass profile is taken 
from the top plot,
while the luminosity profile is taken from the Mosaic CCD data
of RZ04. A galaxy without a dark matter halo, in which mass traces
light, would show a flat line in this plot.}
\label{fig:mass_profile}
\end{figure}

The mass distribution resulting from the Jeans equation calculation,
using the GC velocity dispersion profile found here and the 
best-fitting R$^{1/4}$-law profile for the GC surface density
distribution from RZ04,
is shown in the top panel of Figure \ref{fig:mass_profile}. We have
imposed a cutoff in the GC surface density profile at 15\arcmin\
(projected) radius, consistent with RZ04, but
we have checked that cutoff radii of 30\arcmin\ or 60\arcmin\ only
change the mass profile by a few percent between 1$-$9.5\arcmin\ 
(deprojected) radius.
In Figure \ref{fig:mass_profile}, the upper and lower lines are the 
$1\sigma$ upper and lower limits on the mass (top) and (M/L)$_V$
(bottom). These are based on the bootstrapped uncertainties of the
velocity dispersion profile, which dominate over any uncertainty in
the density profile of the GC system. The dashed line at small radii
is the mass inferred from the rotation curve fit given by Kormendy \&
Westpfahl (1989), based on their analysis of a variety of data in the
literature.  The mass profiles inferred from the isotropic Jeans
equation for our GC sample and from the rotation seen in the disk are
in excellent agreement. This agreement suggests that deviations from
the assumptions in the Jeans equation analysis, such as the known
ellipticity of the isophotes in the central regions (e.g. Burkhead
1986) and/or any anisotropy, produce only modest changes in the mass
distribution which are less than the given uncertainties. 
In Appendix A, we tabulate L$_V$(r) and the 1$\sigma$ limits for M(r) and
(M/L)$_V$(r).

The mass profile inferred from the GC kinematics in 
Figure \ref{fig:mass_profile} continues to rise roughly linearly
with radius from the inner regions to at least 
$7'$ ($\sim$ 20 kpc), after which 
it begins to level off. This naturally follows from the dispersion 
profile in Figure \ref{fig:dispersion} which is approximately constant 
to about this radius accounting for projection, and then falls in an 
almost Keplerian way 
at larger radii. The roughly linear rise in the enclosed
mass from about $\sim 0.5$ to $\sim 4$ $R_e$ ($R_e$ = 105\arcsec\
or 5 kpc) is solid evidence of a dark matter halo
around M104. To show this more explicitly, in the bottom panel of
Figure \ref{fig:mass_profile} we plot the V-band mass-to-light ratio
against deprojected radius r for M104.
This is derived directly from the mass profile in the top panel
and the deprojected
V-band luminosity profile from Kormendy \& Westpfahl (1989). This
luminosity profile agrees to within 5\% with our analysis of 
the Mosaic CCD data of RZ04, and is also in agreement with other
previously published work (e.g. Burkhead 1986).

The radial dependence of the mass-to-light ratio of M104
plotted in Figure \ref{fig:mass_profile} has two major features:
a rise in the $(M/L)_V$ ratio from a value of $\sim$ 4 at
$r = 0.7'$ to a value of $\sim$ 17 at $r = 7'$
($\sim$ 20 kpc, or 4 R$_e$), and a flattening in M/L beyond
7\arcmin. We discuss each of these in turn.
There seems to be no way to account for the rise in M/L
except for a significant dark matter halo around M104. There is no
evidence for a change in the stellar population of the galaxy
over this radial range, as the colors show little or no change over
this region, and in any case, it is hard to imagine how to
produce such a large M/L change in a red spheroidal population.
The increase in M/L from r = $0.7'- 7'$ is also too large to be
easily explained by an orbital structure which changes with radius. 
Moreover, the comparison with the rotation curve 
supports isotropy for the GC orbits in the inner half, and
it would seem contrived for the orbits to suddenly become more
tangential just at the point where the rotation curve measurements end.

Thus the data strongly support the presence of a dark matter halo in
M104. One can also use this analysis to place some broad constraints
on the properties of this dark matter halo.
We can use our results to estimate the fraction of dark matter at one
$R_e$ (105\arcsec, or 5 kpc). To do this, we note that even the 
upper limit for the
mass-to-light ratio at $0.7'$ of $M/L_V = 5.3$ is not larger than
the $M/L_V$ of models of stellar populations with red colours like 
the M104 halo. Specifically, this $M/L_V$ is
less than expected from models with a Salpeter IMF down to 
$0.1 M_{\odot}$, and within the range of models with fewer 
low-mass stars than Salpeter, such as the IMF of Chabrier (2003).
We discuss
this further below, but here note that this suggests that the
mass at 0.7\arcmin\ has little dark matter contribution. Setting the
$M/L_V$ value at $r \sim 0.7'$ as the stellar value and using the
absence of a significant color gradient as evidence that this applies
out to large radii, we can then compute the mass fraction of dark
matter required to account for the mass at one $R_e$. When we do
this, we find a dark matter fraction at one R$_e$ of 19\%. 
This fraction would increase somewhat if the mass at 0.7\arcmin\
was not completely dominated by stars. We also
note that depending on the exact choice of stellar M/L, the halo may
be isothermal or somewhat shallower in density profile over the radial
range of $0.7'-7'$, corresponding approximately to $2-20$ kpc or
$0.4-4$ R$_e$. This is consistent with findings of close to
isothermal mass profiles from strong lensing (Rusin et al. 2003, Treu
\& Koopmans 2004) over a radial range encompassed by this study.

Our data also allow us
to note that at $2 R_e$ the dark matter fraction of the
mass is about $49\%$ and at $5 R_e$ it has risen to about $75\%$. 
M104 thus provides a strong argument against proposals in
which early-type galaxies are either strongly baryon
or dark matter dominated around one R$_e$: instead
we find a mix. While a wide range of
possibilities has been stated at various times in the
literature, this result is in good agreement with the
kinematic studies of the integrated light of elliptical
galaxies (e.g. Gerhard et al. 2001), as well as the 
strong lensing studies of a few
more distant galaxies noted above (Treu \& Koopmans 2004).

We note that the $(M/L)_V$ inferred in our analysis
at $\sim$ 0.7\arcmin\ (0.4 R$_e$) is in the range $3-5$. We can compare
this mass-to-light ratio with those obtained from stellar population
models consistent with the observed colours of the halo
of M104. For $(B-V) \simeq 0.95$, found by both Burkhead (1986)
and our own Mosaic data, stellar population models with a Salpeter
IMF give significantly higher mass-to-light ratios than observed.
M104's disk does not affect the result significantly,
as the disk makes up only about $10\%$ of the light; thus, even
if one assumes the disk contributes no mass, the correction
to the M/L ratio is not significant. The same result is
found when using any of the current stellar population
models (e.g. Vazdekis et al. 1996, Bruzual \& Charlot 2003,
Maraston 2005). The conclusion then is that the mass-to-light
ratio at $\sim 0.4 R_e$ indicated by both GC velocities
and the galaxy rotation curve requires fewer low-mass stars 
well below the turnoff than given by a Salpeter IMF extended
to $0.1 M_{\odot}$.
Similar conclusions
have been drawn from other studies of the central regions
of elliptical galaxies (e.g. Cappellari et al.\ 2006),
and the suggestion of an IMF with fewer low-mass stars
than Salpeter has a long history in 
both elliptical (e.g. Larson 1986, Zepf \& Silk 1996) and
spiral (Bell \& de Jong 2001) galaxies.

Our kinematic data extend well beyond the $\sim$ 20 kpc above,
with our most distant point at $\sim$ 60 kpc. Although our
dispersion profile becomes too uncertain at these largest
radii to reasonably constrain the mass profile, our data do
show a drop in the velocity dispersion outside 
$\sim$ 20 kpc (Figure \ref{fig:dispersion}). Our best fit is that
this drop in dispersion is steep enough to be nearly Keplerian,
and thus the simplest interpretation of the enclosed mass profile
is that there is little mass beyond this radius. This can be
seen in the mass profile at large radii in Figure \ref{fig:mass_profile}.
This apparent flattening of the mass profile beyond 20 kpc radius is not 
affected by the radial extent of the GC surface density profile adopted; 
as noted above, the mass profile only changes by a few percent for cutoff
radii of 15\arcmin, 30\arcmin, or 60\arcmin.
However, there are viable alternatives to the conclusion that the dark
matter halo does not extend much beyond 20 kpc. The most 
obvious issue is whether the assumption of isotropic orbits
holds at these larger radii or whether the orbits become
more radial, which would reconcile the low dispersion estimate
with a higher mass. Another possibility is that the low velocity
dispersion is simply a several sigma fluctuation.
Both of these can be addressed
with significantly more GC velocities at larger radii from the 
center of M104. Fortunately, there are GCs to be observed at these
distances (e.g.\ RZ04), so the study of the dark matter
halo of M104 can be pushed to yet larger radii.

% Authors may indicate to the editorial staff where they would like
% figures and tables to be placed in the manuscript.  This is done with
% either the \placefigure{KEY} or \placetable{KEY} commands.  These
% commands require \label{KEY} commands to be placed appropriately with
% corresponding table and figure captions.  When the manuscript is
% printed a short note is printed on the page where the figure or table
% is to go.  These commands are ignored in the aaspp4 and aas2pp4 styles.

%\placetable{tbl-3}
%\placefigure{fig1}

\section{Conclusions}
\label{conclusions}

We have obtained new velocities for 62 globular clusters (GCs) in M104,
56 of these with 2dF on the AAT and 6 with Hydra on WIYN. Combined
with previous data from Bridges et al. (1997) and Larsen et al.
(2002), we have a total sample of 108 GC velocities, making
M104 one of the few galaxies with more than 100 GC
velocities. Our data extend out to 20\arcmin\ ($\sim$ 60 kpc)
in galactocentric radius, allowing us to study M104 well out
into its halo. Our main conclusions are as follows:

\begin{enumerate}

\item We see no evidence for rotation in the GC system. This is true
for the entire GC system, and also for subsets split by colour and
galactocentric radius. This lack of rotation is particularly
interesting, because there
is rotation of 300$-$350 km/s in the stellar and gas disk (van der Marel
et al. 1994; Kormendy \& Westpfahl 1989), and cosmological simulations
of galaxy formation predict significant amounts of angular momentum in
early-type galaxies. A possible resolution is that galaxy mergers  
transport angular momentum to large radii during the formation
of early-type galaxies. There is a suggestion that the GCs counter-rotate
with respect to the stars and gas at large radius.
We plan to obtain GC velocities to even
larger radii in M104 to better quantify the rotation in the GC system.

\item The GC velocity dispersion is 200$-$250 km/s within $\sim$ 3\arcmin\
of the galaxy center, but then drops to $\sim$ 150 km/s at 10\arcmin,
and possibly flattens at larger radii.

\item Using our GC dispersion profile and the GC spatial profile
from RZ04,
together with the Jeans equation assuming isotropy and spherical
symmetry, we determine the mass distribution of M104 out to
$\sim$ 10\arcmin\ radius ($\sim$ 30 kpc). There is excellent agreement
between the mass profile derived from the GCs and the profile 
inferred from the rotation
curve fit of Kormendy \& Westpfahl (1989) out to 3\arcmin\ radius,
supporting our assumption of isotropy in the GC system.

Our mass profile rises roughly linearly with radius out to about 
7\arcmin\ ($\sim$ 20 kpc), beyond which it levels off. 
The M/L$_V$ increases from $\sim$ 4 at
r = 0.7\arcmin\ to $\sim$ 17 at r=7\arcmin\ (20 kpc, or 4 R$_e$). It
seems difficult to attribute the increase in M/L to changes in 
either the stellar population or GC anisotropy with radius. {\it The data
thus strongly support the presence of a dark matter halo in M104}. We
find that dark matter contributes $\sim$ 20\% of the total mass within
one effective radius, which is in agreement with studies of the
integrated light of elliptical galaxies, and strong lensing studies of
more distant galaxies.

\item We find a M/L$_V$ of 3$-$5 at 0.4 R$_e$, which is difficult to
explain with stellar population models assuming a Salpeter IMF down to 0.1
M$_\odot$. This M/L$_V$ can be accounted for by an IMF with fewer low
mass stars than Salpeter, which has also been found in many other
environments including the Milky Way and other early-type galaxies.

\item More GC velocities, particularly at large radius, will tighten
our interesting constraints on the
rotation and the velocity dispersion/mass profile of the M104 GC system.
It would be very interesting to carry out a detailed comparison of
GC and planetary nebulae velocities, in order
to test for similarities or differences between the
orbital properties of these two populations.

\end{enumerate}

\acknowledgments

We would like to thank the anonymous referee for a very careful reading
of our paper, and for several suggestions which greatly improved it.
We thank Dr. Arunav Kundu for his help with the astrometry of 
the Mosaic data.
KLR is supported by an NSF Astronomy and Astrophysics Postdoctoral
Fellowship under award AST-0302095. SEZ acknowledges support
from NSF award
AST-0406891. TJB would like to thank Dave Hanes for financial support
during the writing of this paper.

\appendix
\section{Appendix A}

In this Appendix, we present in tabular form the luminosity
and mass profiles plotted in Figure \ref{fig:mass_profile}. 
The enclosed V-band L$_V$(r) profile was determined from the Mosaic images
of M104 discussed in Rhode \& Zepf (2004), supplemented
by an analysis of images from the Hubble Space Telescope in
the very central regions where the Mosaic data were saturated.
The enclosed mass profiles M(r) were derived as discussed in the
text, with the upper and lower M(r) representing the $1\sigma$
confidence limits. (M/L)$_V$(r) is simply M(r)/L(r), with upper
and lower values again representing $1\sigma$ confidence limits.

{\bf The full version of Table 5 is available in the ApJ online 
edition of this paper, or by request from T. Bridges}

\begin{deluxetable}{cccccc}
\tabletypesize{\scriptsize}
%\rotate
\tablecaption{Enclosed luminosity, mass, and mass-to-light ratio 
versus deprojected radius r for M104.
Column 1: radius in arcmin; Column 2: enclosed luminosity in solar units;
Columns 3\&4: lower and upper 1$\sigma$ confidence limits on the enclosed
mass in solar units; 
Columns 5\&6: lower and upper 1$\sigma$ confidence limits on the
(M/L)$_V$.}
\tablehead{
r & L$_V$($<$r) & M$_{low}$($<$r) & M$_{high}$($<$r) &
(M/L)$_{V,low}$(r) & (M/L)$_{V,high}$(r) \\
(arcmin) & (Log(L$_\odot$)) & (Log(M$_\odot$)) & (Log(M$_\odot$)) & & \\
}
%\tableline
\startdata
0.708 & 10.042 & 10.549 & 10.763 & 3.21 & 5.25 \\
0.724 & 10.053 & 10.564 & 10.774 & 3.24 & 5.26 \\
0.740 & 10.064 & 10.577 & 10.787 & 3.26 & 5.29 \\
0.756 & 10.074 & 10.590 & 10.801 & 3.28 & 5.33 \\
0.772 & 10.085 & 10.602 & 10.813 & 3.29 & 5.36 \\
...   & ... & ... & ... &  ... & ... \\
11.924 & 10.755 & 11.578 & 11.822 & 6.66 & 11.67 \\
11.940 & 10.755 & 11.540 & 11.826 & 6.09 & 11.77 \\
11.956 & 10.755 & 11.519 & 11.780 & 5.80 & 10.59 \\
11.972 & 10.755 & 11.500 & 11.754 & 5.56 & 9.97 \\
11.988 & 10.756 & 11.630 & 11.805 & 7.49 & 11.20 \\

\enddata
\end{deluxetable}

\clearpage

%\section{Floating material and so forth}

%\clearpage

% Now comes the reference list.  In this document, we used \cite to call
% out citations, so we must use \bibitem in the reference list, which
% means we use the LaTeX thebibliography environment.  Please note that
% \begin{thebibliography} is followed by a null argument.  If you forget
% this, mayhem ensues, and LaTeX will say "Perhaps a missing item?" when
% you run it.  Do not call us, do not send mail when this happens.  Put
% the silly {} after the \begin{thebibliography}.
%
% Each reference has a \bibitem command to define the citation format
% to be placed in the text (in []) and the symbolic tag used for
% cross referencing (in {}).
%
% See sample1.tex, or the AASTeX guide, for an alternative to the \cite-
% \bibitem command.


\begin{thebibliography}{}

\bibitem[Ashman \& Zepf (1992)]{az92} Ashman, K.M., \& Zepf, S.E. 1992,
\apj, 384, 50

\bibitem[Ashman \& Zepf (1998)]{az98} Ashman, K.M., \& Zepf, S.E. 1998,
Globular Cluster Systems (Cambridge: Cambridge University Press)

\bibitem[Barbier-Brossat et al. (1994)]{barbier94} Barbier-Brossat, M.,
Petit, M., \& Figon, P. 1994, A\&AS, 108, 603

\bibitem[Barbier-Brossat \& Figon (2000)]{barbier00} Barbier-Brossat, M.,
\& Figon, P. 2000, A\&AS, 142, 217

\bibitem[Beers et al. (1990)]{beers90} Beers, T.C., Flynn, K.,
\& Gebhardt, K. 1990, \aj, 100, 32

\bibitem[Bell \& de Jong (2001)]{bell01} Bell, E.F.,
\& de Jong, R.S. 2001, \apj, 550, 212

\bibitem[Bridges \& Hanes (1992)]{tjb92} Bridges, T.J., \& Hanes, D.A.
1992, \aj, 103, 800

\bibitem[Bridges et al.\ (1997)]{tjb97} Bridges, T.J., Ashman, K.M.,
Zepf, S.E., Carter, D., Hanes, D.A., Sharples, R.M., \& Kavelaars, J.J.
1997, MNRAS, 284, 376 [B97]

\bibitem[Bridges et al.\ (2006)]{tjb06} Bridges, T.J., Gebhardt, K.,
Sharples, R., Faifer, F.R., Forte, J.C., Beasley, M.A., Zepf, S.E.,
Forbes, D.A., Hanes, D.A., \& Pierce, M. 2006, MNRAS, in press

\bibitem[Brodie \& Strader (2006)]{brodie06} Brodie, J.P., 
Strader, J. 2006, ARA\&A, in press (astroph/0602601)

\bibitem[Bruzual \& Charlot (2003)]{bruzual03} Bruzual, G., \& 
Charlot, S. 2003, MNRAS, 344, 1000

\bibitem[Burkhead (1986)]{burk86} Burkhead, M.S. 1986, \aj, 91, 777

\bibitem[Cappellari et al. (2006)]{cappellari06} Cappellari, M., 
Bacon, R., Bureau, M., Damen, M.C., Davies, R.L., de Zeeuw, P.T.,
Emsellem, E., Falcon-Barroso, J., Krajnovic, D., Kuntschner, H.,
McDermid, R.M., Peletier, R.F., Sarzi, M., van den Bosch, R.C.E.,
\& van de Ven, G. 2006, MNRAS, 366, 1126

\bibitem[Chabrier (2003)]{chabrier03} Chabrier, G. 2003, ApJ,
586, L133

\bibitem[Cohen (2000)]{cohen00} Cohen, J.G. 2000, \aj, 119, 162

\bibitem[C\^ot\'e et al.\ (2001)]{cote01} C\^ot\'e, P., McLaughlin, D.E.,
Hanes, D.A., Bridges, T.J., Geisler, D., Merritt, D., Hesser, J.E.,
Harris, G.L.H., \& Lee, M.G. 2001, \apj, 559, 828

\bibitem[C\^ot\'e et al.\ (2003)]{cote03} C\^ot\'e, P., McLaughlin, D.E.,
Cohen, J.G., \& Blakeslee, J.P. 2003, \apj, 591, 850

\bibitem[Croom et al. (2005)]{croom05} Croom, S., Saunders, W.,
Heald, R., \& Bailey, J., ``The 2dfdr Data Reduction System Users
Manual, available at www.aao.gov.au/AAO/2df/manual.html

\bibitem[Dekel et al. (2005)]{dekel05} Dekel, A., Stoehr, F.,
Mamon, G.A., Cox, T.J., Novak, G.S., \& Primack, J.R. 2005, Nature,
437, 707

\bibitem[de Vaucouleurs et al. (1991)]{devauc91} de Vaucouleurs, G.,
de Vaucouleurs, A., Corwin, H.G., Buta, R.J., Paturel, G.,
\& Fouque, P. 1991, ``Third Reference Catalogue of Bright Galaxies'',
Springer-Verlag

\bibitem[Faber et al. (1977)]{faber77} Faber, S.M., Balick, B.,
Gallagher, J.S., \& Knapp, G.R. 1977, \apj, 214, 383

\bibitem[Forbes et al.\ (1997)]{fbg97} Forbes, D.A., Brodie, J.P., \&
Grillmair, C.J. 1997, \aj, 113, 1652

\bibitem[Gebhardt \& Kissler-Patig (1999)]{geb99} Gebhardt, K. \&
Kissler-Patig, M. 1999, \aj, 118, 1526

\bibitem[Ford et al. (1996)]{ford96} Ford, H.C., Hui, X., Ciardullo, R.,
Jacoby, G.H., \& Freeman, K.C. 1996, \apj, 458, 455

\bibitem[Gerhard et al. (2001)]{gerhard01} Gerhard, O., Kronawitter, A.,
Saglia, R.P., \& Bender, R. 2001, \aj, 121, 1936

\bibitem[Harris \& van den Bergh (1981)]{harris81} Harris, W.E.,
\& van den Bergh, S. 1981, \aj, 86, 1627

\bibitem[Harris et al. (1984)]{harris84} Harris, W.E., Harris, H.C.,
\& Harris, G.L.H. 1984, \aj, 89, 216

\bibitem[Harris (1996)]{harris96} Harris, W.E. 1996, \aj, 112, 1487

\bibitem[Harris \& van den Bergh (1981)]{hvdb81} Harris, W.E. \& van
den Bergh, S. 1981, \aj, 86, 1627

\bibitem[Hempel et al. (2006)]{hempel06} Hempel, M., Zepf, S.E.,
Kundu, A., Geisler, D., \& Maccarone, T.J. 2006, in preparation

\bibitem[Hernquist \& Bolte (1993)]{hernquist93} Hernquist, L.,
\& Bolte, M. 1993, in ``The Globular Clusters-Galaxy Connection'',
ASP Conference Series \#48, ed. G.H. Smith \& J.P. Brodie, pg. 788

\bibitem[Kent (1988)]{kent88} Kent, S.M. 1988, /aj, 96, 514

\bibitem[Kormendy \& Westpfahl (1989)]{kormendy89}  Kormendy, J., \& 
Westpfahl, D.J. 1989 \apj, 338, 752

\bibitem[Kronawitter et al. (2000)]{kronawitter00} Kronawitter, A.,
Saglia, R.P., Gerhard, O., \& Bender, R. 2000, A\&AS, 144, 53

\bibitem[Kundu \& Whitmore (2001)]{kw01} Kundu, A. \& Whitmore,
B.C. 2001, \aj, 121, 2950

\bibitem[Larsen et al. (2001)]{larsen01} Larsen, S.S., Forbes, D.A.,
\& Brodie, J.P. 2001, MNRAS, 327, 1116

\bibitem[Larsen et al. (2002)]{larsen02} Larsen, S.S., Brodie, J.P.,
Beasley, M.A., \& Forbes, D.A. 2002, \aj, 124, 828 [L02]

\bibitem[Larson (1986)]{larson86} Larson, R.B. 1986, MNRAS, 218, 409

\bibitem[Lewis et al. (2002)]{lewis02} Lewis, I.J., Cannon, R.D.,
Taylor, K., Glazebrook, K., Bailey, J.A., Baldry, I.K.,
Barton, J.R., Bridges, T.J., Dalton, G.B., Farrell, T.J.,
Gray, P.M., Lankshear, A., McCowage, C., Parry, I.R., Sharples, R.M.,
Shortridge, K., Smith, G.A., Stevenson, J., Straede, J.O.,
Waller, L.G., Whittard, J.D., Wilcox, J.K., \& Willis, K.C. 2002,
MNRAS, 333, 279

\bibitem[Malaroda et al. (2001)]{malaroda01} Malaroda, S.,
Levato, H., \& Galliani, S. 2001, VizieR On-line Data Catalog:
III/216

\bibitem[Maraston (2005)]{maraston05} Maraston, C. 2005, MNRAS, 362, 799

\bibitem[Monet et al. (1998)]{monet98} Monet, D., Bird, A.,
Canzian, B., Dahn, C., Guetter, H., Harris, H., Henden, A.,
Levine, S., Luginbuhl, C., Monet, A.K.B., Rhodes, A., Riepe, B.,
Sell, S., Stone, R., Vrba, F., \& Walker, R. 1998, 
``The USNO-A2.0 Catalogue'', VizieR On-Line Data Catalog: I/252,
US Naval Observatory

\bibitem[Peng et al. (2004)]{peng04} Peng, E.W., Ford, H.C.,
\& Freeman, K.C. 2004, \apj, 602, 705

\bibitem[Perrett et al. (2002)]{perrett02} Perrett, K.M.,
Bridges, T.J., Hanes, D.A., Irwin, M.J., Brodie, J.P.,
Carter, D., Huchra, J.P., \& Watson, F.G. 2002, \aj, 123, 2490

\bibitem[Petrov et al. (2006)]{petrov06} Petrov, L., 
Kovalev, Y.Y., Fomalont, E.B., \& Gordon, D. 2006, \aj, 131, 1872

\bibitem[Rhode \& Zepf (2001)]{rhode01} Rhode, K.L. \& Zepf,
S.E. 2001, \aj, 121, 210

\bibitem[Rhode \& Zepf (2004)]{rhode04} Rhode, K.L. \& Zepf,
S.E. 2004, \aj, 127, 302 [RZ04]

\bibitem[Richtler et al. (2004)]{richtler04} Richtler, T.,
Dirsch, B., Gebhardt, K., Geisler, D., Hilker, M., Alonso, M.V.,
Forte, J.C, Grebel, E.K., Infante, L., Larsen, S., Minniti, D.,
\& Rejkuba, M. 2004, /aj, 127, 2094

\bibitem[Roberts \& Haynes (1994)]{roberts94} Roberts, M.S.,
\& Haynes, M.P. 1994, ARA\&A, 32, 115

\bibitem[Rubin et al. (1978)]{rubin78} Rubin, V.C., Thonnard, N.,
\& Ford, W.K. 1978, \apj, 225, 107

\bibitem[Rusin et al. (2003)]{rusin03} Rusin, D., Kochanek, C.S.,
\& Keeton, C.R. 2003, \apj, 595, 29

\bibitem[Schlegel, Finkbeiner, \& Davis (1998)]{schlegel98}
Schlegel, D., Finkbeiner, D., \& Davis, M. 1998, ApJ, 500, 525

\bibitem[Spitler et al. (2006)]{spitler06} Spitler, L.R.,
Larsen, S.S., Strader, J., Brodie, J.P., Forbes, D.A., \& Beasley, M.A.
2006, \aj, in press (astroph/0606337)

\bibitem[Tonry et al. (2001)]{tonry01} Tonry, J.L., Dressler, A.,
Blakeslee, J.P., Ajhar, E.A., Fletcher, A.B., Luppino, G.A.,
Metzger, M.R., \& Moore, C.B. 2001, \apj, 546, 681

\bibitem[Treu \& Koopmans (2004)]{treu04} Treu, T., \& Koopmans, L.V.E.
2004, \apj, 611, 739

\bibitem[van der Marel et al.\ (1994)]{vdM94} van der marel, R.P.,
Rix, H.-W., Carter, D., Franx, M., White, S.D.M., \& de Zeeuw, T.
1994, \mnras, 268, 521

\bibitem[Vazdekis et al. (1996)]{vazdekis96} Vazdekis, A., Casuso, E.,
Peletier, R.F., \& Beckman, J.E. 1996, \apjs, 106, 307

\bibitem[Vitvitska et al. (2002)]{vitvitska02} Vitvitska, M.,
Klypin, A.A., Kravtsov, A.V., Wechsler, R.H., Primack, J.R.,
\& Bullock, J.S. 2002, \apj, 581, 799

\bibitem[Zepf et al.\ (2000)]{zepf00} Zepf, S.E., Beasley, M.A.,
  Bridges, T.J., Hanes, D.A., Sharples, R.M., Ashman, K.M., \&
  Geisler, D. 2000, \aj, 120, 2928

\bibitem[Zepf \& Silk (1996)]{zs96} Zepf, S.E. \& Silk, J. 1996, \apj,
466, 114

\end{thebibliography}
\end{document}